\documentclass[%
 aip,
 amsmath,amssymb,
]{revtex4-1}

\usepackage{graphicx}
\usepackage{dcolumn}
\usepackage{bm}
\usepackage[version=4]{mhchem}
\usepackage[utf8]{inputenc}
\usepackage[T1]{fontenc}
\usepackage{mathptmx}
\usepackage{etoolbox}

\usepackage{lineno}

\makeatletter
\def\@email#1#2{%
 \endgroup
 \patchcmd{\titleblock@produce}
  {\frontmatter@RRAPformat}
  {\frontmatter@RRAPformat{\produce@RRAP{*#1\href{mailto:#2}{#2}}}\frontmatter@RRAPformat}
  {}{}
}%
\makeatother
\begin{document} 


\title[]{Graphics Processing Unit/Artificial Neural Network-accelerated large-eddy simulation of turbulent combustion: Application to swirling premixed flames}
\author{Min Zhang}
\affiliation{ 
College of Engineering, Peking University, Beijing, 100871, P.R. China
}%
 \affiliation{AI for Science Institute (AISI), Beijing, 100080, China.}
\author{Runze Mao}%
\affiliation{ 
College of Engineering, Peking University, Beijing, 100871, P.R. China
}%

 \affiliation{AI for Science Institute (AISI), Beijing, 100080, China.}

\author{Han Li}%
\affiliation{ 
College of Engineering, Peking University, Beijing, 100871, P.R. China
}%

 \affiliation{AI for Science Institute (AISI), Beijing, 100080, China.}

\author{Zhenhua An}%

\affiliation{ 
Department of Mechanical Engineering and Science, Kyoto University, Kyoto daigaku-Katsura, Nishikyo-ku, Kyoto 615–8540, Japan
}%

\author{Zhi X. Chen}
 \email{chenzhi@pku.edu.cn.}

\affiliation{ 
College of Engineering, Peking University, Beijing, 100871, P.R. China
}%
 
  \affiliation{AI for Science Institute (AISI), Beijing, 100080, China.}


\date{\today}

\begin{abstract}
Within the scope of reacting flow simulations, the real-time direct integration  (DI) of stiff ordinary differential equations (ODE) for the computation of chemical kinetics stands as the primary demand on computational resources. Meanwhile, as the number of transport equations that need to be solved increases, the computational cost grows more substantially, particularly for those combustion models involving direct coupling of chemistry and flow such as the transported probability density function model. 
In the current study, an integrated Graphics Processing Unit-Artificial Neural Network (GPU-ANN) framework is introduced to comply with heavy computational costs while maintaining high fidelity. Within this framework, a GPU-based solver is employed to solve partial differential equations and compute thermal and transport properties, and an ANN is utilized to replace the calculation of reaction rates.
Large eddy simulations of two swirling flames provide a robust validation, affirming and extending the GPU-ANN approach's applicability to challenging scenarios.
The simulation results demonstrate a strong correlation in the macro flame structure and statistical characteristics between the GPU-ANN approach and the traditional Central Processing Unit~(CPU)-based solver with DI. This comparison indicates that the GPU-ANN approach is capable of attaining the same degree of precision as the conventional CPU-DI solver, even in more complex scenarios. In addition, the overall speed-up factor for the GPU-ANN approach is over two orders of magnitude. This study establishes the potential groundwork for widespread application of the proposed GPU-ANN approach in combustion simulations, addressing various and complex scenarios based on detailed chemistry, while significantly reducing computational costs.
\end{abstract}

\maketitle


\section{\label{sec:level1}Introduction}


Large-eddy simulation (LES) with a good capability to model the interaction of turbulence and chemistry is commonly used for simulating canonical flames. However, there are very few studies reported concerning high-fidelity LES of turbulent flames in real industrial cases, mainly due to heavy computational costs. In the realm of reacting flow simulations, it is widely acknowledged that the computation of chemical kinetics through real-time direct integration (DI) of stiff ordinary differential equations~(ODE) is the predominant consumer of computational resources. To illustrate, in cases where the reaction involves more than 50 species, the chemical kinetics component can account for approximately 90\,\% of the total computational effort~\cite{avdic2017}. 

In addressing the aforementioned challenge, significant endeavors have been undertaken to reduce the computational demands of processing chemical kinetics by streamlining the real-time solution of ODE. One of the most important options for decreasing the computational burden associated with the DI of ODE is to utilize various forms of tabulation methods that leverage a pre-computed database of reaction source terms~\cite{Peters2001, chen1989}. Among these approaches, flamelet-generated mamifolds~(FGM) is a popular technique that uses several variables~(such as mixture fraction and progress variable) to look up the pre-computed table~\cite{van2000}. The successful implementation of the FGM in the turbulent flame can be found in~\cite{zhang2021, baik2022}. It should be noted that more look-up variables are required to tackle complicated conditions. This leads to a significant increase in memory requirements. To minimize memory usage and enable the use of large-scale mechanisms, two innovative methods have been introduced: intrinsic low-dimensional manifolds (ILDM)~\cite{Maas1992} and in-situ adaptive tabulation (ISAT)~\cite{pope1997}, with the latter involving the dynamic updating of chemistry tabulation throughout the simulation. Nevertheless, tabulation techniques ease the computational load while potentially compromising the accuracy of physical fidelity. For instance, species that undergo slower chemical reactions, such as acetylene~(\ce{C2H2}) and nitrogen dioxide~(\ce{NO}), require a longer duration to adjust to changes in the dissipation rate, thereby resulting in an underestimation of their concentration levels~\cite{ketelheun2011}. In addition, the underestimations of \ce{C2H2} and \ce{NO} consequently lead to prediction errors of soot and nitrogen oxides~(\ce{NOx}). As a result, the aforementioned tabulation methods may not be an effective option in certain scenarios.  

It is well-known that employing finite rate chemistry (FRC) in combustion modeling typically results in more precise reaction processes without limiting the scenarios. Recently, there's been a fast-paced growth in the field of artificial intelligence (AI), mainly in the area called machine learning (ML). This growth has opened up fresh views on speeding up detailed FRC while still maintaining high accuracy. A common approach to accelerate FRC is to apply a neural network as a substitute for the DI of the stiff ODE~\cite{Zhou2022, Ihme2022}.
Christo et al.~\cite{christo1995} employed a single artificial neural network (ANN) to tabulate simple mechanisms, specifically the 3-step and 4-step chemical systems involved in combustion, showing the potential of ANN to substitute the ODE integration. In the prior study by Brown et al.~\cite{brown2021}, parallel residual networks (ResNets) were employed to address the stiff ODE dictated by the chemically reacting flows. This was executed using a simplified hydrogen~\ce{H2}-air reaction model that encompassed 8 species and 18 reactions~\cite{petersen1999}. Recently, several feature reduction techniques coupled with ANN have been employed when large data sets were involved.     
For instance, Blasco et al.~\cite{blasco2000, blasco1998} partitioned the composition space into multiple subdomains within thermochemical space (such as mixture fraction or temperature) using the automatic partitioning technique known as self-organizing map (SOM). Each subdomain was then fitted with an individual ANN. In the study~\cite{parente2009, abdelwahid2023}, principal component analysis (PCA) was combined with an ANN to enhance the size-reduction capabilities of PCA and address the inefficiencies of the statistic-based nonlinear regression model in fitting highly nonlinear datasets. Additionally, Castellanos et al.~\cite{castellanos2023} proposed time-lag autoencoders, which enhance the reconstruction of the thermochemical process's temporal progression. However, this study did not employ an ANN model as a substitute for the ODE integrator. More recently, Goswami et al.~\cite{Goswami2024} have successfully integrated an autoencoder into the DeepONet frameworks, demonstrating exceptional accuracy for species that evolve over considerably shorter time scales.

In practice, the primary challenge when using the ANN to accelerate the direct integration of ODEs lies in ANN that can generalize across a wide range of realistic combustion scenarios. To achieve such generalization capabilities, the crucial step is to sample training data comprehensively. An et al.~\cite{an2020} trained an ANN using the training data collected from a Reynolds-averaged-Navier-Stokes~(RANS) simulation and then applied it to the LES case. High-precision results, in terms of ignition quenching and mass fraction prediction, can be achieved in comparison to a conventional ODE integrator. 
Weng et al.~\cite{wenga2023} sampled data from the LES case and extended the concept of Fourier Neural Operator (FNO) to learn stiff chemical kinetics.
It is apparent that sampling training data from a specified RANS/LES case limits the generalization capabilities of the trained ANN.  
In contrast to sample training data from the RANS case, a multi-scale sampling method has been introduced for sampling data in several studies~\cite{zhang2022, xu2024}.
For instance, Zhang et al.~\cite{zhang2022} used the multi-scale sampling method to collect training data across the full composition space, demonstrating strong generalization capability for a range of flame configurations. However, training costs rise significantly with more than 5 million samples and upwards of 1.6 million model parameters.  
Recently, Readshaw~\cite{readshaw2023} and Ding et al.~\cite{ding2021} proposed a hybrid method, known as HFRD-MMLP (Hybrid Flamelet/Random Data and Multiple Multilayer Perceptrons), to simulate the Cambridge stratified flame. The same technique was applied to various fuels (dimethyl ether, and blends of methane~(\ce{CH4}) and \ce{H2}) and flame configurations (Sandia flame D and the Sydney bluff-body flame), demonstrating exceptional accuracy relative to the direct integration of ODE and robust generalization capability~\cite{ding2022, liu2024}. It is widely recognized that as the number of transport equations to be solved increases, the computational cost becomes more significant, thereby diminishing the overall acceleration factor despite the use of ANN to expedite the integration of ODE. This phenomenon is illustrated in the study~\cite{readshaw2023}, where the reaction source term is reduced by a factor of fourteen, yet the overall acceleration factor is only about four.

Graphics processing units (GPUs), originally engineered for the highly parallel task of rendering graphics, can greatly enhance the speed of computational fluid dynamics (CFD) simulations~\cite{perez2018, bielawski2023, Esclapez2023, henry2023}. Integrating an ANN model into a GPU-based CFD solver is expected to further reduce the aforementioned bottleneck associated with computational costs while maintaining high fidelity. To achieve this, a GPU-AI approach is proposed and employed to simulate the Xi'an Jiao Tong University~(XJTU) swirling flame~\cite{an2021} and Cambridge high-swirling stratified flame~\cite{sweeney2012}. The LES of these two swirling flames serves as a robust validation to affirm and extend the proposed GPU-AI approach to complicated turbulent conditions.  

The remainder of this paper is structured as follows: the integrated GPU-AI framework including GPU acceleration, training data, and ANN model is briefly introduced in Section~\ref{sec2}. The evaluation of the integrated GPU-ANN approach is performed by comparing the macro flame structure of the XJTU swirling flame and detailed statistical characteristics of the Cambridge Stratified flame in Section~~\ref{sec3} and Section~\ref{sec4}, respectively. The computational saving is analyzed in Section~\ref{sec5}. Lastly, concluding remarks are provided in Section~\ref{sec6}.

\section{Integrated GPU-AI framework}
\label{sec2}

\begin{figure}
\includegraphics[width=192pt]{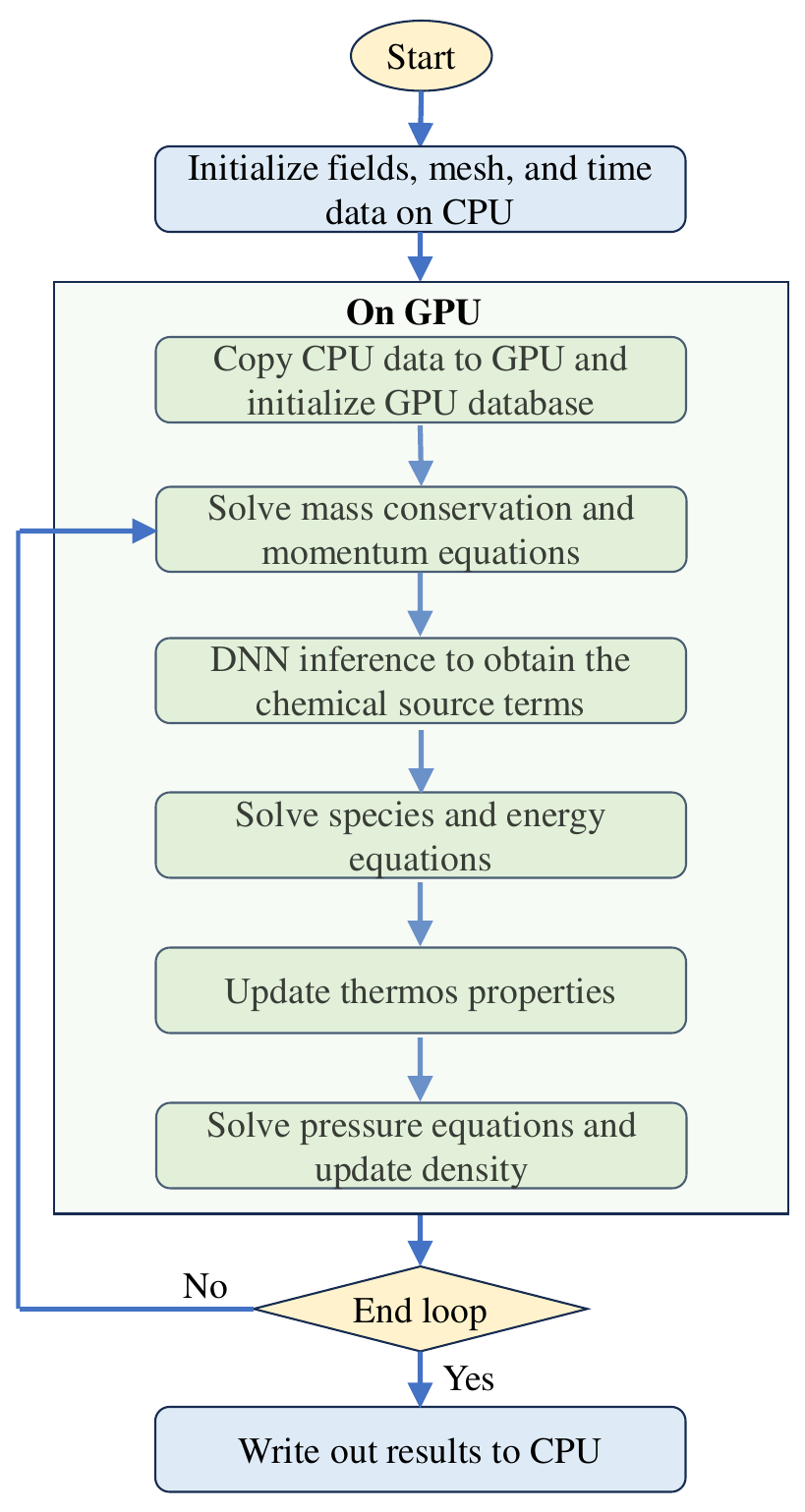}
\caption{\label{fig:1} Flow chart of integrated GPU-AI framework.}
\end{figure}

In the authors' previous work~\cite{mao2023deep}, the solvers are Central Processing Unit~(CPU)-based in the developed open-source code, DeepFlame~\cite{mao2023deep}, which enhances its performance by integrating the robust fluid dynamics functionalities of OpenFOAM~\cite{Weller1998}, the specialized thermochemical analysis capabilities of Cantera~\cite{goodwin2002}, and the sophisticated machine learning algorithms provided by PyTorch~\cite{paszke2019}. It is important to highlight that DeepFlame is not a distinct methodology; instead, it serves as an open-source platform that allows researchers to implement their own ANN models, acceleration techniques, and simulation scenarios. As an example, the present study has successfully incorporated an integrated GPU-AI framework. Details on this GPU-AI framework are described as follows,

\subsection{\label{sec:level2}GPU acceleration framework}

Figure~\ref{fig:1} shows the framework of GPU acceleration for dfLowMachFoam, which is one of the solvers for DeepFlame open-source code. 
As illustrated in Fig.~\ref{fig:1}, only the initializing data at the first step and writing out the simulation results are carried out on CPU to avoid the CPU-GPU memory copy overhead. The computational process executed on GPU mainly can be categorized into three parts:
\begin{itemize}
\item \textbf{Solve Partial Differential Equations (PDEs).} Fully implicit pressure-based PDEs are solved for mass, momentum, species, energy, and pressure using the finite volume method (FVM). The sparse matrix, which represents a linear system post-discretization in the implicit FVM, has its storage format transformed from the lower-diagonal-upper (LDU) to the compressed sparse row (CSR) format within the current GPU framework. 
The implicit discretization is executed with compute unified device architecture (CUDA) kernel functions to guarantee precise management of threads and memory. Subsequently, the Algebraic Multigrid Solver~(AmgX) library~\cite{naumov2015} produced by NVIDIA is utilized to solve the linear system.

\item \textbf{Infer chemical source terms with ANN.} The conventional DI of ODE is substituted by ANN inference. The ANN-related operations are implemented based on the libTorch library~\cite{collobert2002}. To enhance computational efficiency during the inference procedure, the model adopts the half-precision floating-point format. Furthermore, a dynamic inference batch size is employed to mitigate the GPU memory footprint.

\item \textbf{Update thermal and transport properties.} The update operations for thermal and transport properties, involving Newton's method and high-order temperature polynomials, are conducted through CUDA functions.
\end{itemize}

Additionally, for direct GPU communication, the NVIDIA collective communication library (NCCL) is employed instead of the conventional message-passing interface (MPI) approach to achieve multi-processor parallelism. Moreover, various optimizations have been performed to increase computational performance and minimize the GPU memory footprint. The relevant code implementation can be found on our GitHub site~\cite{deepflame2024}.

\begin{figure*}
\includegraphics[width=480pt]{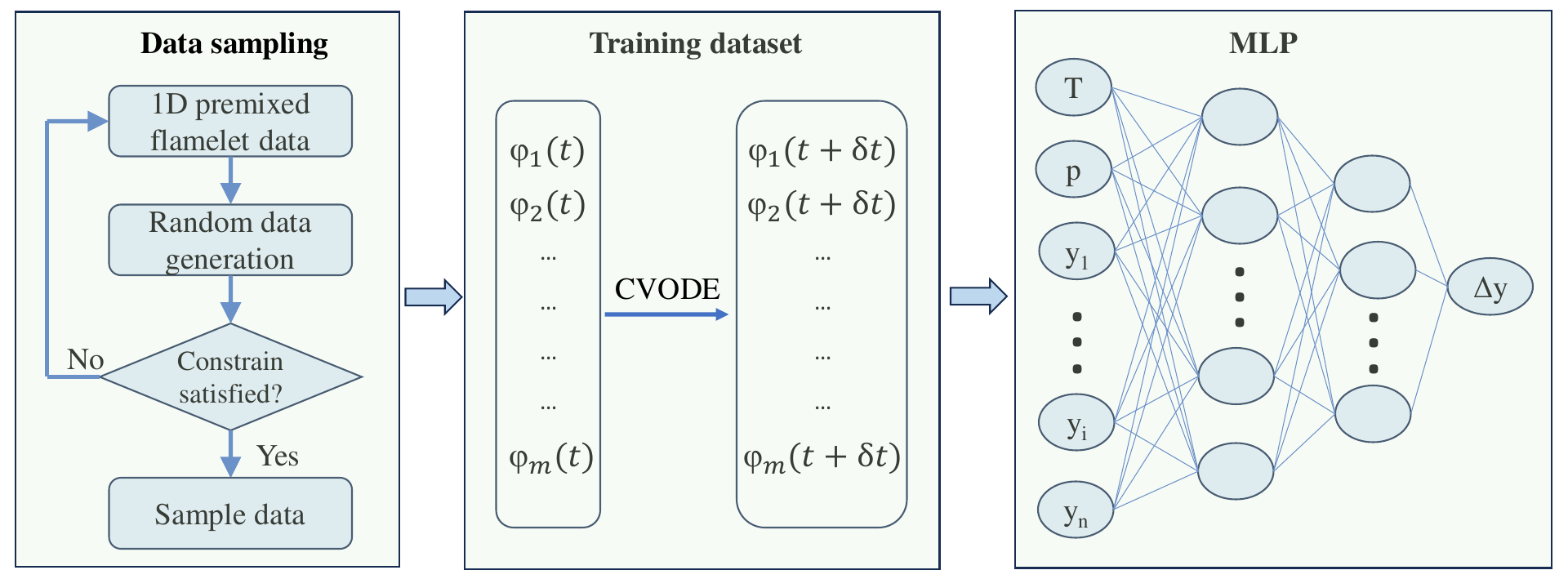}
\caption{\label{fig:2} Scheme of the process of training data and ANN model generation.}
\end{figure*}

\subsection{\label{sec:level3}Training data and DNN model generation}

The instantaneous reaction rates of a given mechanism, determined by local species concentration~($y_{i}$), temperature~($T$), and pressure~($p$), can be solved by the following systems of ODEs:
\begin{equation}
\Dot{\omega}_{k} = \frac{dy_{i}}{dt} = \varphi (y_{i},T,p)
\label{omega}
\end{equation}
Given a specific initial composition, the numerical integration of Eq.~\ref{omega} over a time step, $\delta t$, produces an input-output pair as follows:
\begin{equation}
\varphi(t) 	\Rightarrow \varphi(t + \delta t)
\label{input}
\end{equation}
The core principle behind substituting ANN for reaction rate calculations lies in the fact that the procedure, referenced in Eq.~\ref{input}, can be approximated through a non-linear optimization process, widely known as training. 
The hybrid flamelet/random data~(HFRD) methodology proposed by Readshaw and Ding et al.~\cite{readshaw2023, ding2021} is employed in the present study. Only a brief training data generation procedure and its new developments are presented here.
Fig.~\ref{fig:2} illustrates a brief description of the procedures for employing ANN methodology in the present study, which consists of the following three steps: 
\begin{itemize}
\item \textbf{Sample and generate random data from flamelet data.} This step primarily involves sampling data from flamelet simulations and generating random data points based on the sampled flamelet data. Initial data is sourced from one-dimensional (1-D) premixed flame simulations which are carried out with the DRM19 chemical mechanism consisting of 20 species and 85 reactions~\cite{Kazakov}. Subsequently, the dataset derived from the 1-D simulations provides the basis for random data generation. For every composition within the flamelet dataset, a new random composition is generated through the following procedures:
\begin{equation}
T' = T + \alpha \gamma (T_{max} - T_{min}),
\label{Trand}
\end{equation}
\begin{equation}
p' = p + \beta \gamma (p_{max} - p_{min}),
\label{Prand}
\end{equation}
\begin{equation}
y'_{N_{2}} = y_{N_{2}} + \alpha \gamma (y_{N_{2},max} - y_{N_{2},min}),
\label{N2rand}
\end{equation}
\begin{equation}
y'_{j} = y_{j}^{(1+ \theta \gamma )},
\label{Yrand}
\end{equation}
where the parameters are set as $\alpha$, $\beta$, and $\theta$ are set to 0.125, 10.0, and 0.1, respectively. $\gamma$ is random number within [-1,1] with uniform distribution. Moreover, the randomly generated data, following the mentioned method, should adhere to various specified constraints. Initially, the mass fractions of species must collectively sum to unity. Additionally, the molar element ratio and equivalence ratio of each composition state in the random dataset must possess appropriate values. Specifically, $\text{H/C} \in [2.65, 4.67]$, $\text{O/N} \in [0.254, 0.320]$, and equivalence ratio $\in [0.20, 1.15]$ are used in the present study. Fig.~\ref{fig:3} shows a comparison between flamelet data, random data, random data with interpolation, and data collected from the LES simulation of Cambridge stratified flame. It is clear that the flamelet data after random generation almost covers the LES data. However, the mixing of extremely lean reactants with high-temperature products, which primarily occurs due to entrainment from large eddy structures, cannot be accounted for, as shown in Fig.~\ref{fig:3}(b). As a result, linear interpolation is used for all species, bridging the equilibrium compositions of the lowest equivalence ratio and the ambient air compositions, as illustrated in Fig.~\ref{fig:3}(c). 
\begin{figure*}
\includegraphics[width=480pt]{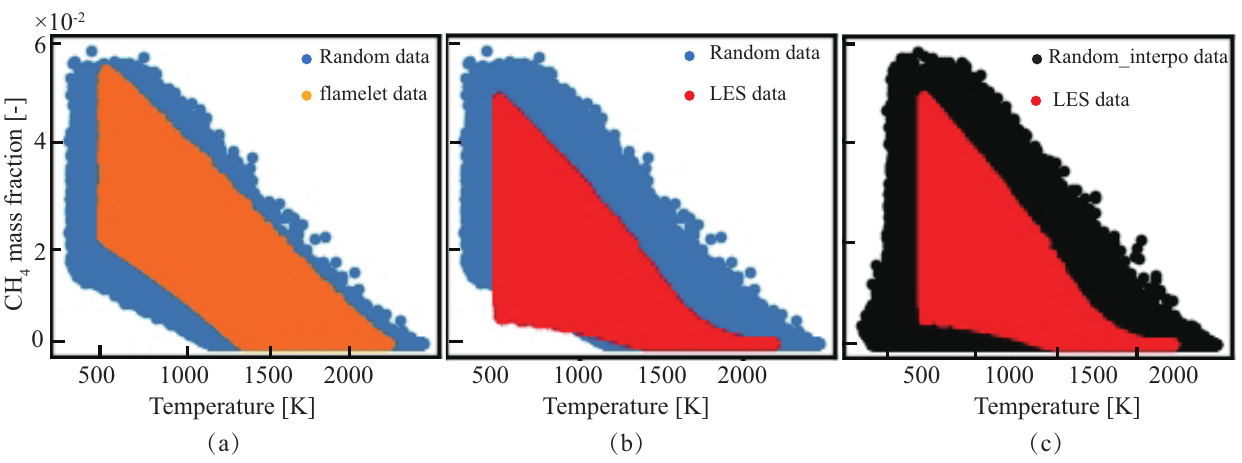}
\caption{\label{fig:3}Comparison of scatter plots between the flamelet dataset, random dataset, random data after interpolation, and LES data collected from Cambridge swirling flame in $T$-\ce{CH4} space.}
\end{figure*}

\item \textbf{Generate training data.}
After generating the random dataset, each composition within it is processed using the CVODE integrator~\cite{brown1989} for a time step of $1\,\mathrm{\mu s}$ to obtain the target output. In this study, a total of 518,000 input-output pairs are produced to form the final training dataset.
\item \textbf{ANN training using multilayer perceptrons~(MLP).} 
The MLP approach is employed in the present study. According to Eq.~\ref{omega}, the input layers of the MLP can be represented as $\varphi(t)=\{T(t), p(t), \mathcal{F}(\boldsymbol{Y}(t))\}$. The output layers represent the variation in the mass fractions of species across a specified time interval, represented as $\boldsymbol{u}^*(t)=\boldsymbol{Y}(t+\Delta t) - \boldsymbol{Y}(t)$. 
Since the different species exhibit disparate orders of magnitude in thermochemical phase space throughout the evolution of the chemical system, the Box-Cox transformation (BCT), proposed by Box and Cox (1964)~\cite{Box1964} and introduced for pre-process of training data in~\cite{zhang2022}, is employed here to represent multi-scale data by O(1) quantity and avoid the singularity arising from the log transformation when the data approaches zero. 
\begin{equation}
f(x) =  \begin{cases}
            \frac{x^{\lambda -1}}{\lambda} \\
      \text{log}(x) 
         \end{cases}
\label{bct}
\end{equation}
where $\lambda = 0.1$ is adopted in the present study. The mass fraction of the chemical species within [0,1] is mapped to [$-1/ \lambda$, 0] after the BCT. Training and prediction are individually performed for each species' mass fraction to ensure high accuracy. The architecture of each MLP similar to~\cite{zhang2022} includes hidden layers consisting of 1600, 800, and 400 perceptrons, respectively. The network utilizes the Gaussian error linear unit (GELU) as its activation function and hyperparameter optimization is carried out using the Adam algorithm. Training the ANN involves evaluating a loss function that compares its predictions against the training data, which represents the numerically integrated solution of the complete chemical system:
\begin{equation}
\begin{split}
    \mathcal{L}_{data}&=\frac{1}{N} \sum_{i=1}^N \left(\boldsymbol{Y}^*_i(t+\Delta t) - \boldsymbol{Y}_i(t+\Delta t) \right)^{2}
\end{split}
\label{loss1}
\end{equation}
Since the full thermochemical space obeys the conservation of the species mass fractions during the reaction as the solution vector evolves, the loss corresponding to this physical constraint is expressed as follows:
\begin{equation}
\begin{split}
    \mathcal{L}_{mass}&=\left \vert  \sum_{i=1}^N \boldsymbol{Y}^*_i(t+\Delta t) -1   \right \vert
\end{split}
\label{loss2}
\end{equation}  
With the consideration of the heat release rate~($\text{HRR}$), the additional loss is described as follows:
\begin{equation}
\begin{split}
    \mathcal{L}_{HRR}&= \left(\boldsymbol{HRR}^*(t+\Delta t) - \boldsymbol{HRR}(t+\Delta t) \right)^{2}
\end{split}
\label{loss3}
\end{equation}
The total loss function for the ANN takes this form:
\begin{equation}
\begin{split}
    \mathcal{L}_{tot}&= W_{1}\mathcal{L}_{data} + W_{2}\mathcal{L}_{mass} + W_{3}\mathcal{L}_{HRR} 
\end{split}
\label{lossTot}
\end{equation}
where $W_{1}$, $W_{2}$, and $W_{3}$ are the weights for each loss term.
\end{itemize}

\begin{figure}
\includegraphics[width=250pt]{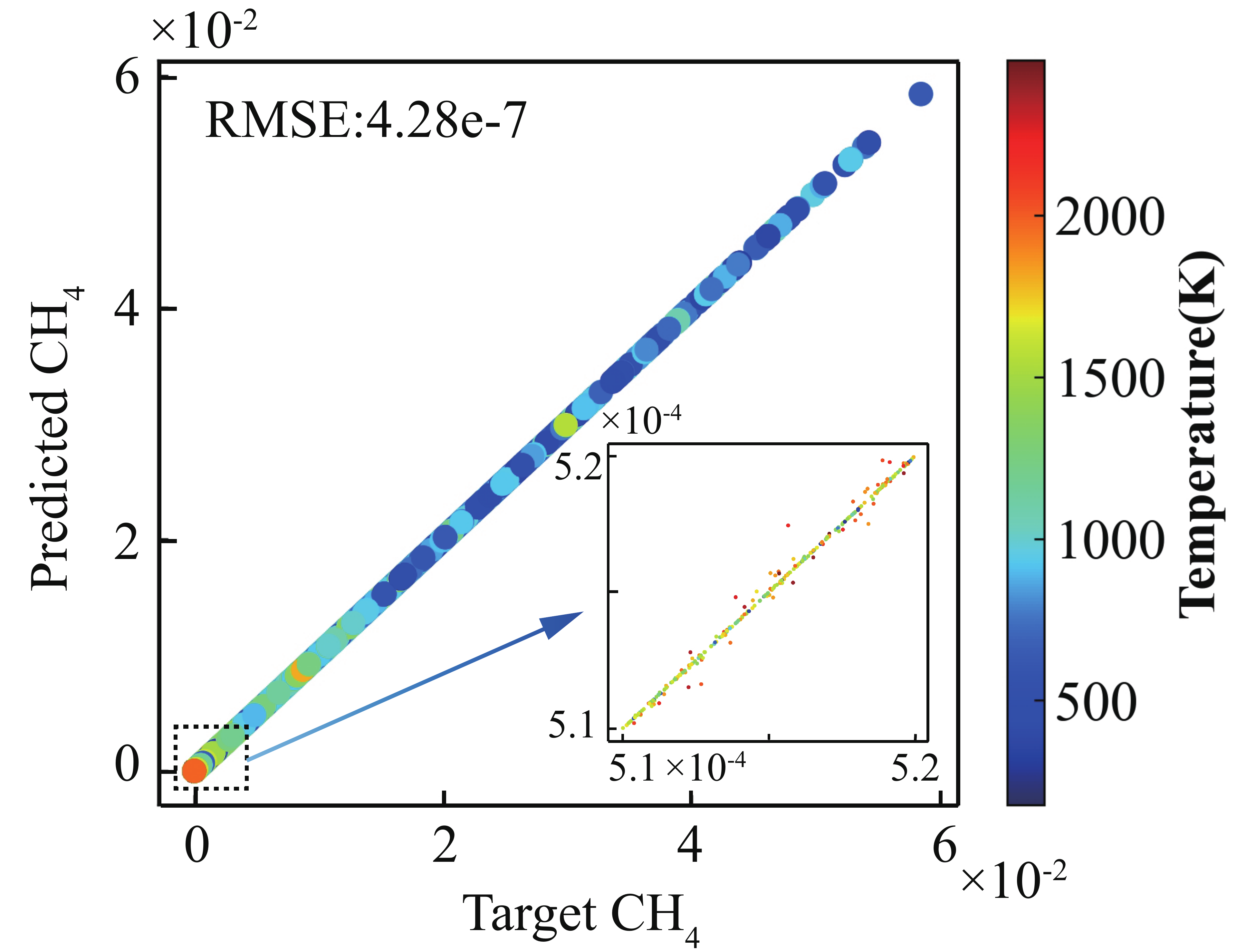}
\caption{\label{fig:4} MLP predictions for \ce{CH4} colored by the temperature of the sample.}
\end{figure}

The prediction performance of the trained ANN is preliminarily assessed using a different random dataset, generated by replicating all the above procedures. A commendable agreement between the predicted and target values is illustrated in Fig~\ref{fig:4} where root-mean-square-error~(RMSE) is included. Meanwhile, a comparable agreement in laminar flame speed under different equivalence ratios is observed (not shown here) between the trained ANN and direct integration of ODE. All these establish the applicability of the trained ANN for subsequent assessments in the swirling flames. To demonstrate that the trained ANN is not case-specific, simulations of the XJTU swirl flame and the Cambridge swirling stratified premixed flame will be presented.

\section{Application to XJTU swirling flame}
\label{sec3}

\subsection{Configuration of XJTU swirl combustor and numerical methods}

\begin{figure*}
\includegraphics[width=420pt]{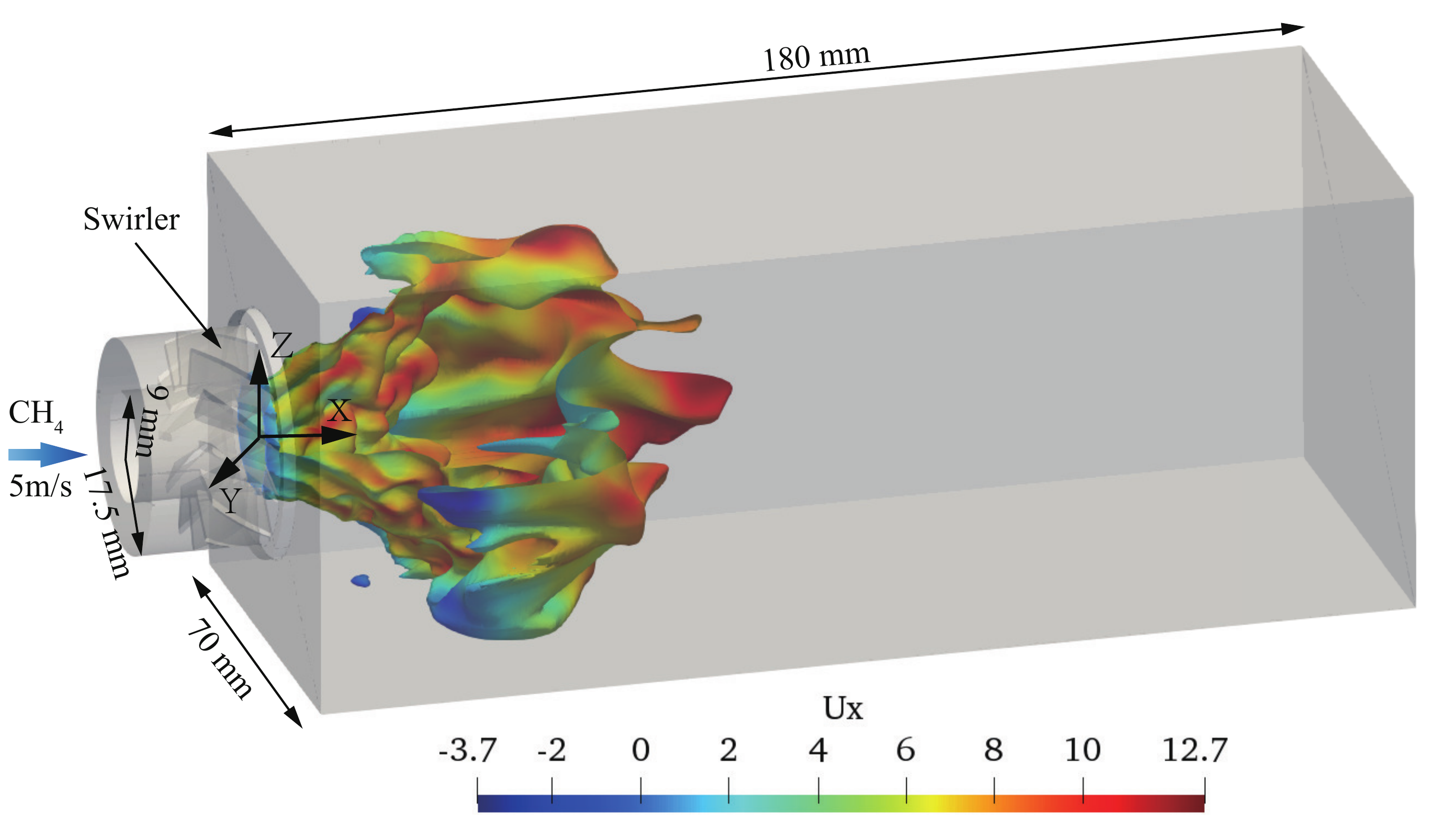}
\caption{\label{fig:5} Computational domain of XJTU burner. The iso-surface of flame temperature of $2000\,\mathrm{K}$ colored by axial velocity.} 
\end{figure*}

\begin{figure*}
\includegraphics[width=480pt]{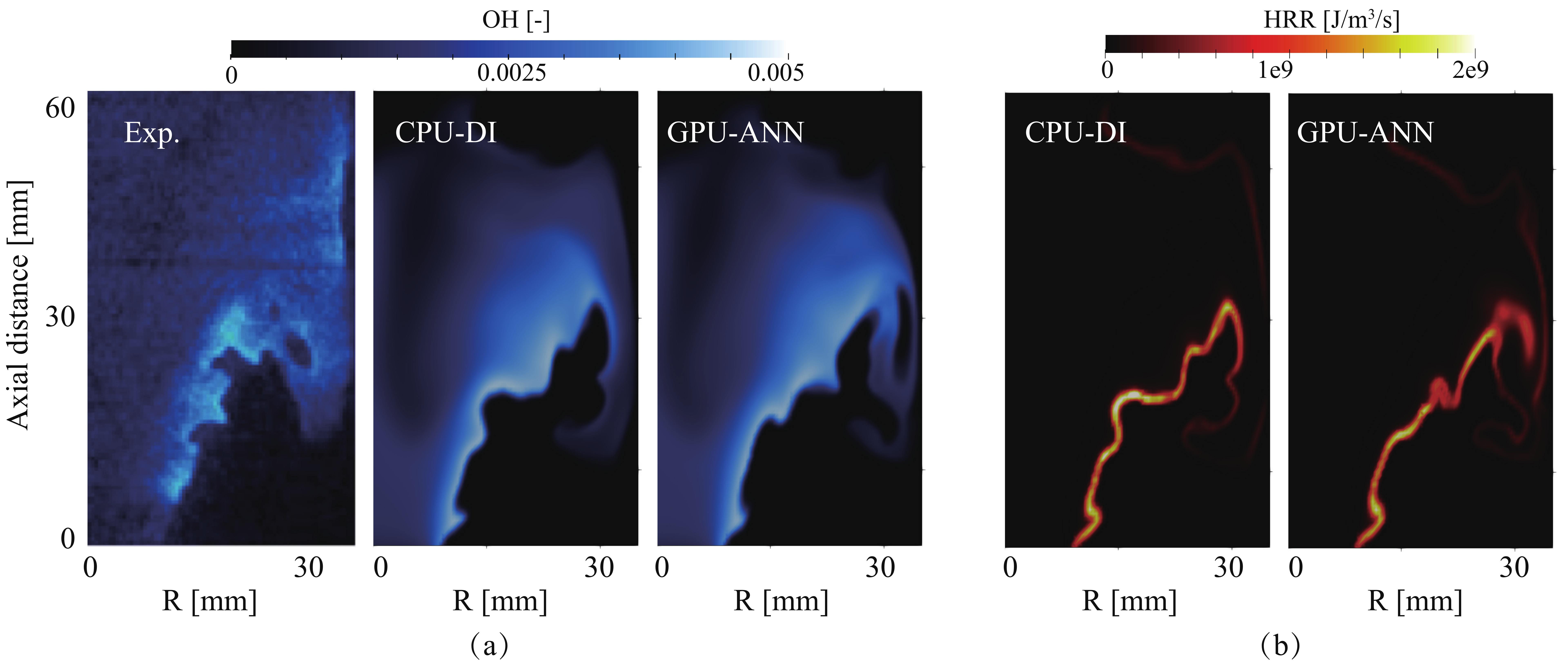}
\caption{\label{fig:6}Comparison of (a) measured OH and simulated OH using the CPU-DI and GPU-ANN approaches as well as (b) heat release rate obtained from the CPU-DI and GPU-ANN approaches.}
\end{figure*} 

Figure~\ref{fig:5} shows the configuration of the XJTU swirl combustor. The combustor is composed of three main parts: a premixing chamber, a 12-vane swirler set at a 45-degree angle, and a quartz liner. Methane~(\ce{CH4}) and air undergo thorough mixing within the premixing chamber. The swirler with 12 vanes and 45 vane angle is used to create a swirling flow to stabilize the flame. The quartz liner with the dimensions of $70 \times 70 \times 180\,\mathrm{mm}$ is used to provide both flame visibility and access for laser diagnostic. In this study, a stoichiometrically premixed methane-air mixture, with an average velocity of $5\,\mathrm{m/s}$, is channeled through the swirler. This process produces a highly swirling flow with a swirl number of 0.71. The detailed structure of this combustor can be found in the studies~\cite{Zhang2021Emiss, zhang2021blow}.  

The computational domain corresponds to the experimental setup, as illustrated in Fig.~\ref{fig:5}. A nonuniform grid with a total of 5.0 million cells is employed. The mesh refinement is specifically applied to the flame region with cell size less than $0.4\,\mathrm{mm}$.
The mesh resolution employed here has been verified by referenced studies~\cite{zhang2021blow, zhang2021regulation, an2021} to be adequate for achieving a high resolution in the flame region. In the current study, the gas-phase flow field is resolved by employing Favre-filtered compressible Navier-Stokes equations, which are coupled with a Smagorinsky model for sub-grid-scale stress. The interaction between turbulence and combustion chemistry is modeled using the partially stirred reactor~(PaSR) combustion model. Within this model, the characteristic chemical time is determined by the slowest species formation rate~\cite{evans2019}, while the mixing time scale is derived from the geometric mean of the Taylor and Kolmogorov scales~\cite{chomiak1996}.
It is important to note that the ANN model only substitutes the direct integration of source terms within the PaSR model, rather than replacing the PaSR model entirely.
The momentum is discretized in space with second-order schemes.  All walls utilize no-slip boundary conditions to replicate physical realism. At the fuel inlet, we introduce pseudo-turbulent fluctuations via a sophisticated synthetic eddy method for turbulence generation~\cite{kornev2007}. The integral length scale and fluctuation intensity are consistent with those documented in referenced works~\cite{Proch2017}.

\subsection{Results and discussion for XJTU swirling premixed flame}

In this study, simulations are carried out using two distinct methodologies: the CPU-DI and GPU-ANN. The CPU-DI method utilizes a CPU-based solver for the resolution of flow and scalar equations, integrating reaction sources through the CVODE solver. Conversely, the GPU-ANN method employs a GPU-based solver, with a trained ANN predicting reaction rates. 
Note that the CPU/GPU-based solvers share identical governing equations and numerical modeling.

In this section, the macro flame structure, such as flame height, combustion intensity, and the distribution of radical species, is first compared to evaluate the predictive performance of the GPU-ANN approach. 
The height of the flame is a critical attribute in characterizing the overall structure of a flame. In line with Singh's definition~\cite{Singh2012}, the height of the flame corresponds to the edge of the reaction zone, as clearly evidenced by the hydroxyl radical~(OH) distribution. 
Fig.~\ref{fig:6}(a) illustrates the comparison of OH distribution between the experimental measurements and the predictions from the GPU-ANN and CPU-DI approaches. It is evident that both approaches predict similar flame height as the experimental measurement. 
In addition, both simulation results capture the sharp increase in OH intensity near the flame surface region, indicating that the GPU-ANN approach is capable of identifying the reaction location in a premixed flame. 
Considering that the OH intensity does not necessarily imply the combustion intensity, the heat release rates (HRR) obtained from the GPU-ANN and CPU-DI approaches are then compared in Fig.~\ref{fig:6}(b). A similar distribution of the HRR indicates the combustion intensity is well captured using the GPU-ANN approach.

\begin{figure*}
\includegraphics[width=470pt]{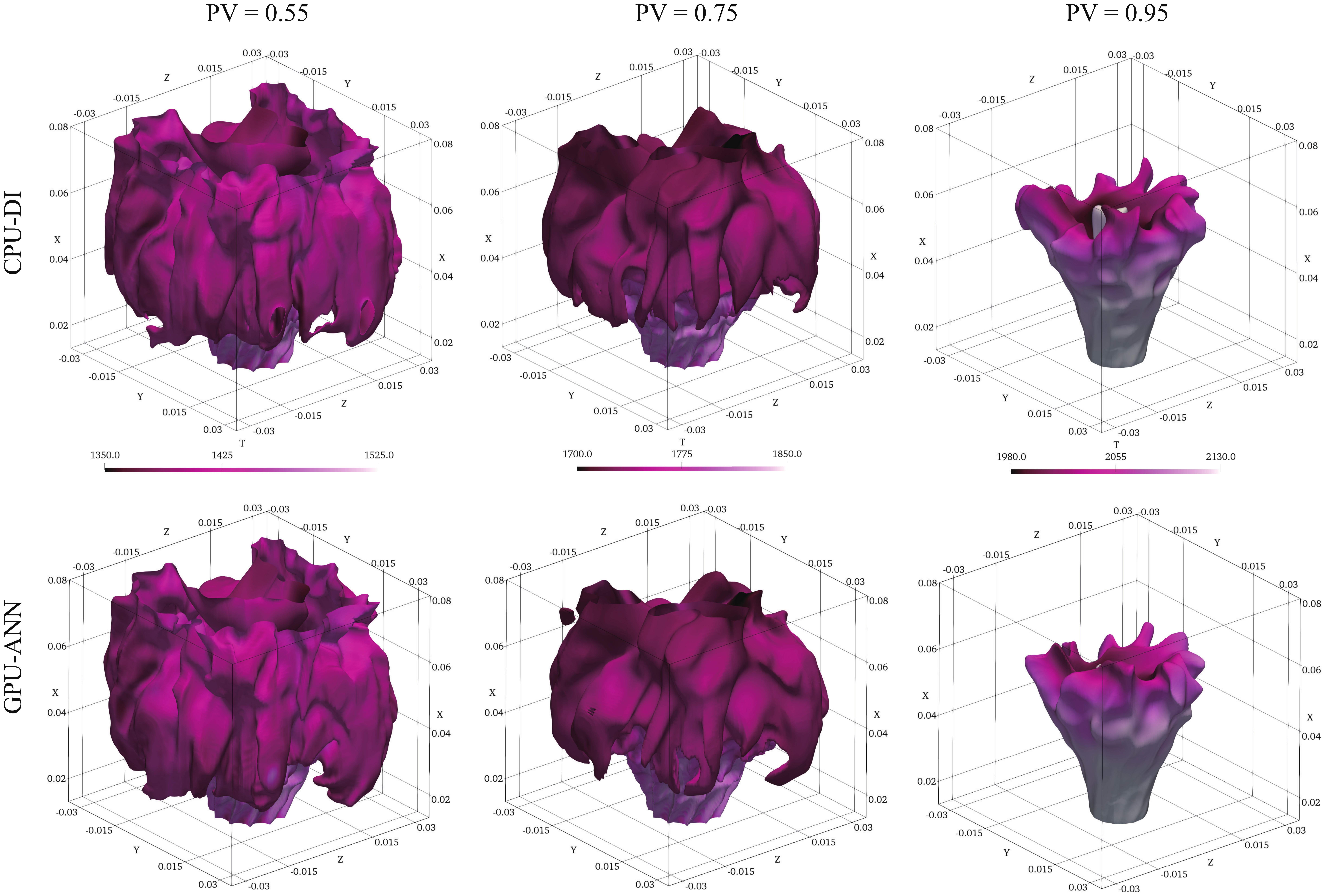}
\caption{\label{fig:7} Comparison of iso-surface of progress variables of 0.55, 0.75, and 0.95 at $12\,\mathrm{ms}$ between the CPU-DI and GPU-DNN approaches.}
\end{figure*}

\begin{figure*}
\includegraphics[width=440pt]{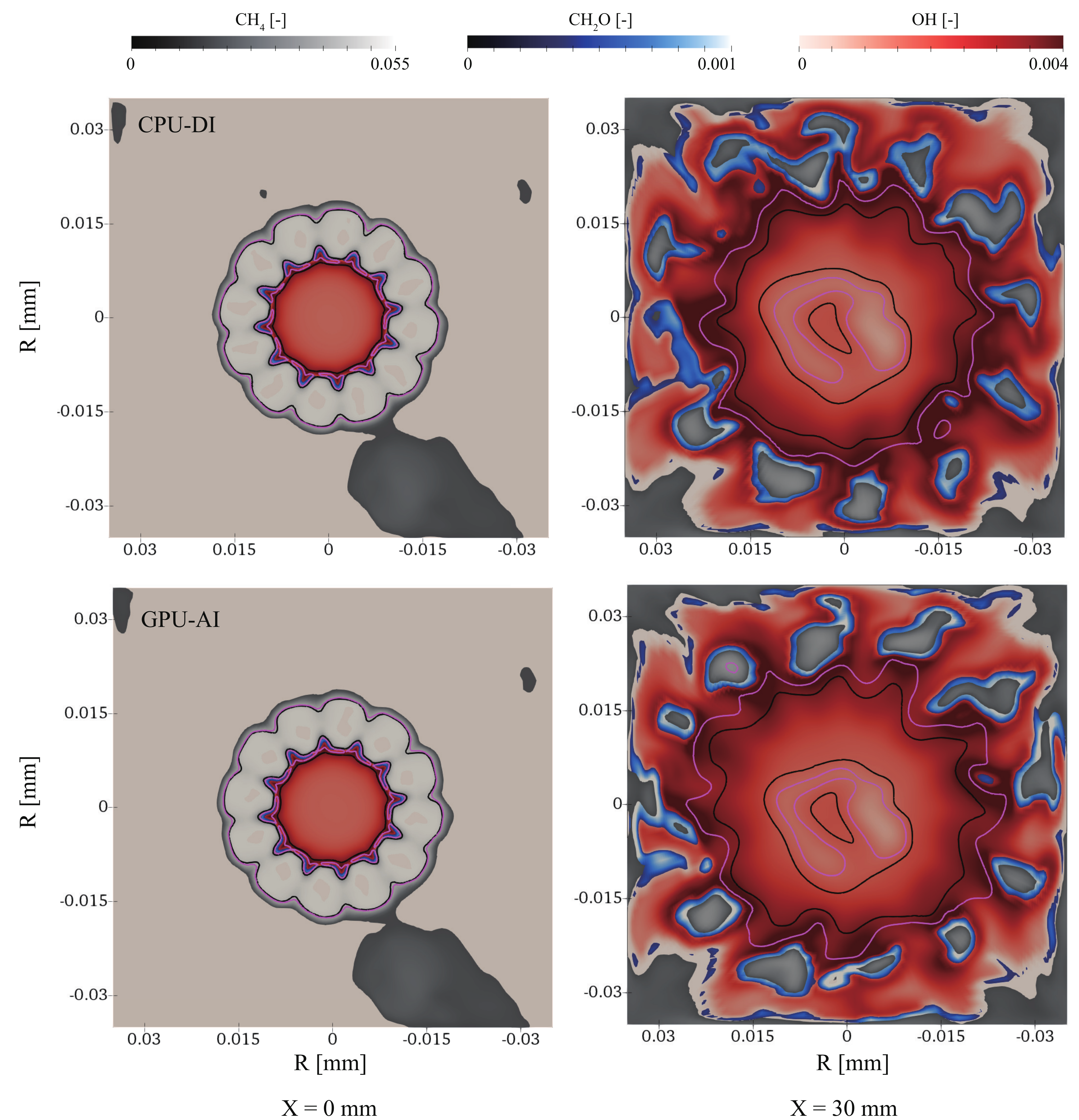}
\caption{\label{fig:8} Comparison of the distributions of CH4-OH-CH2O at $12\,\mathrm{ms}$ for different axial locations of $0\,\mathrm{mm}$ and $30\,\mathrm{mm}$ in the CPU-DI and GPU-ANN cases. The magenta and black lines represent the equivalence ratio of 0.85 and 0.90. The range of iso-surface of \ce{CH4}, \ce{CH2O}, and \ce{OH} are [0.02,~0.055], [$0.4 \times 10^{-3}$, $1.2\times 10^{-3}$], and [$0$, $4\times 10^{-3}$], respectively.}
\end{figure*}

A further evaluation of the GPU-ANN approach is to compare the macro flame structure in terms of the distribution of temperature and key radical species. The three-dimensional~(3-D) topology of premixed flame is presented in Fig.~\ref{fig:7} where the iso-surface of different progress variables~(PV) is colored by the flame temperature. The topology structures of the flame under different PVs appear similar between the CPU-DI and GPU-DNN approaches. Additionally, the range of flame temperature at each PV state remains the same for both approaches. This indicates that the GPU-ANN approach is capable of reproducing the evolution of the flame with respect to the PV.
Fig.~\ref{fig:8} shows the distributions of formaldehyde~\ce{CH2O}-OH-\ce{CH4} at axial locations of $0\,\mathrm{mm}$ and $30\,\mathrm{mm}$ for the CPU-DI and GPU-DNN cases. At $\text{X}= 0\,\mathrm{mm}$, in both the CPU-DI and GPU-AI scenarios, the findings reveal a uniform distribution of the wrinkled flame along the circumferential direction, marked by twelve convex regions that align with the number of vanes. These regions exhibit a lower equivalence ratio, pointing to accelerated fuel consumption attributable to stretch effects~\cite{gaucherand2023}. As the axial distance increases to $30\,\mathrm{mm}$, it is apparent that \ce{CH2O} is encompassed by \ce{OH}, while \ce{CH4} is encompassed by \ce{CH2O}. The highest OH concentration is predominantly located at the equivalence ratio of 0.85. Within the center of the flame, a lower equivalence ratio is observed, which is likely a result of entrainment bringing the downstream mixture into the recirculation zone. All these observations are consistent across both the CPU-DI and GPU-AI scenarios. This consistency provides solid evidence that the GPU-ANN approach is capable of predicting the macro flame structures as effectively as the traditional CPU-DI approach.

\section{Application to Cambridge high-swirling stratified flame}
\label{sec4}

\begin{table}
\caption{\label{tab:table1}Main parameters for Cambridge burner SWB7. }
\begin{ruledtabular}
\begin{tabular}{cccccc}
\text{$U_{o}$}($\mathrm{m/s}$)&\text{$U_{i}$}($\mathrm{m/s}$)&\text{$U_{co}$}($\mathrm{m/s}$)&\mbox{$\phi_{i}$}&\mbox{$\phi_{o}$}&\mbox{Swirl}(\%)\\
\hline
8.3&18.7& 0.4 & 1.0 & 0.5 & 0.79\\
\end{tabular}
\end{ruledtabular}
\end{table}

The proposed GPU-AI approach is qualitatively evaluated in premixed swirling flame by comparing the macro flame structure in Section~\ref{sec3}. In this section, this approach will be applied to Cambridge swirling stratified premixed flame, designed by Sweeney et al.~\cite{sweeney2012, sweeney2012struc}, to further assess its performance with robust statistical analysis of scalars. Initially, both mean and root mean square (RMS) data are time-averaged over a duration approximately equivalent to three complete flow-through times, based on the inner jet velocity and an axial extent of $40\,\mathrm{mm}$. An additional averaging is performed in the azimuthal direction. Furthermore, the analysis of scatters in temperature-species space is performed.

\subsection{Configuration of Cambridge swirling stratified premixed flame and numerical methods}

\begin{figure}
\includegraphics[width=250pt]{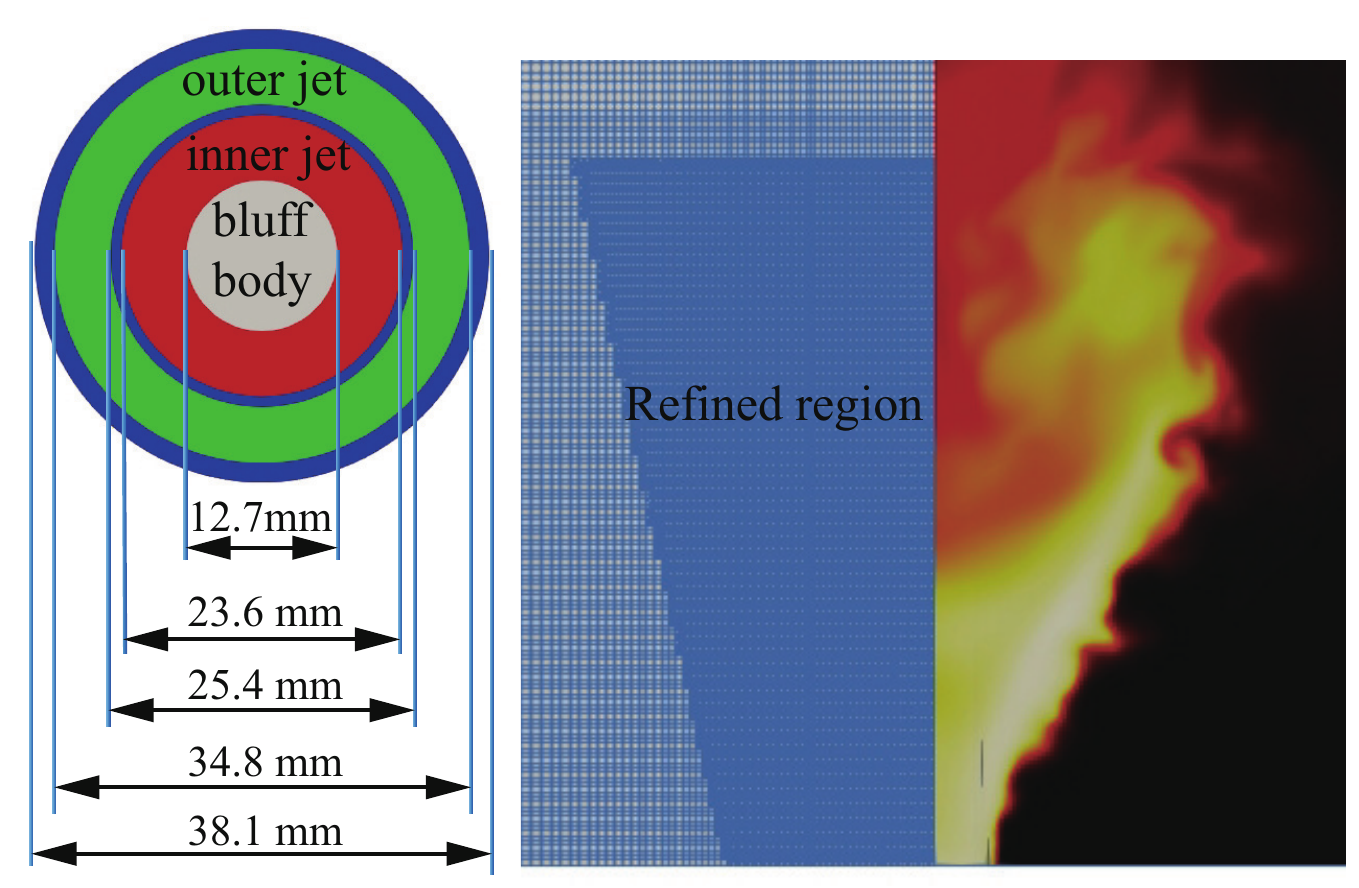}
\caption{\label{fig:9} Dimension of the Cambridge burner and mesh configuration used in the simulation.}
\end{figure}

Compared to the study~\cite{readshaw2023}, a higher swirl ratio case SWB7 is simulated here. Fig.~\ref{fig:9} shows the scheme view of the Cambridge burner.  
The burner incorporates a central bluff body accompanied by dual annular fuel jets. Airflow is facilitated at the burner's exterior. The external fuel jet can provide varied swirling flows, characterized by swirl numbers, which denote the ratio of averaged tangential velocity to averaged axial velocity. The major parameters of the case SWB7 are presented in Table~\ref{tab:table1}. Detailed experimental setup and measurement methods are available in~\cite{sweeney2012struc}.   

The computational domain is a cylindrical domain with the dimensions of $300\,\mathrm{mm} \times 100\,\mathrm{mm} \times 2\pi$ in axial, radial, and azimuthal directions, respectively. As shown in Fig.~\ref{fig:6}, a refined mesh with cell size less than $0.25\,\mathrm{mm}$ is applied to cover the flame region, which can provide enough resolution  
based on the mesh size used in~\cite{zhang2021, qian2022}. The numerical methods and turbulence inlet boundary conditions employed in this case are the same as those used in Section~\ref{sec3}.

\subsection{Results and discussion for Cambridge swirling stratified premixed flame}

\begin{figure*}
\includegraphics[width=440pt]{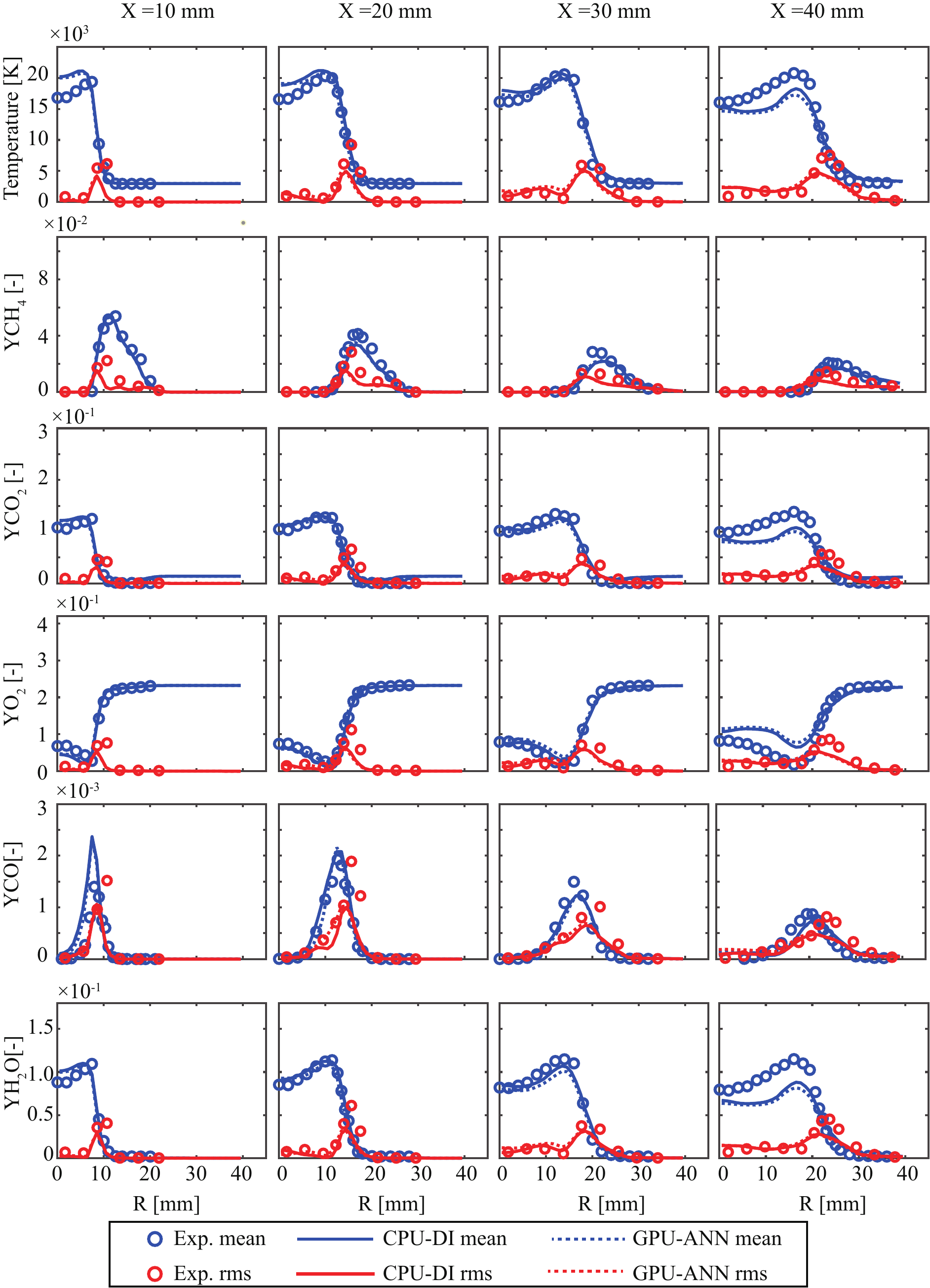}
\caption{\label{fig:10}Radial profiles of mean and RMS for temperature and major species mass fraction at four different axial locations in the experiment, CPU-DI, and GPU-ANN cases.}
\end{figure*}

\begin{figure*}
\includegraphics[width=440pt]{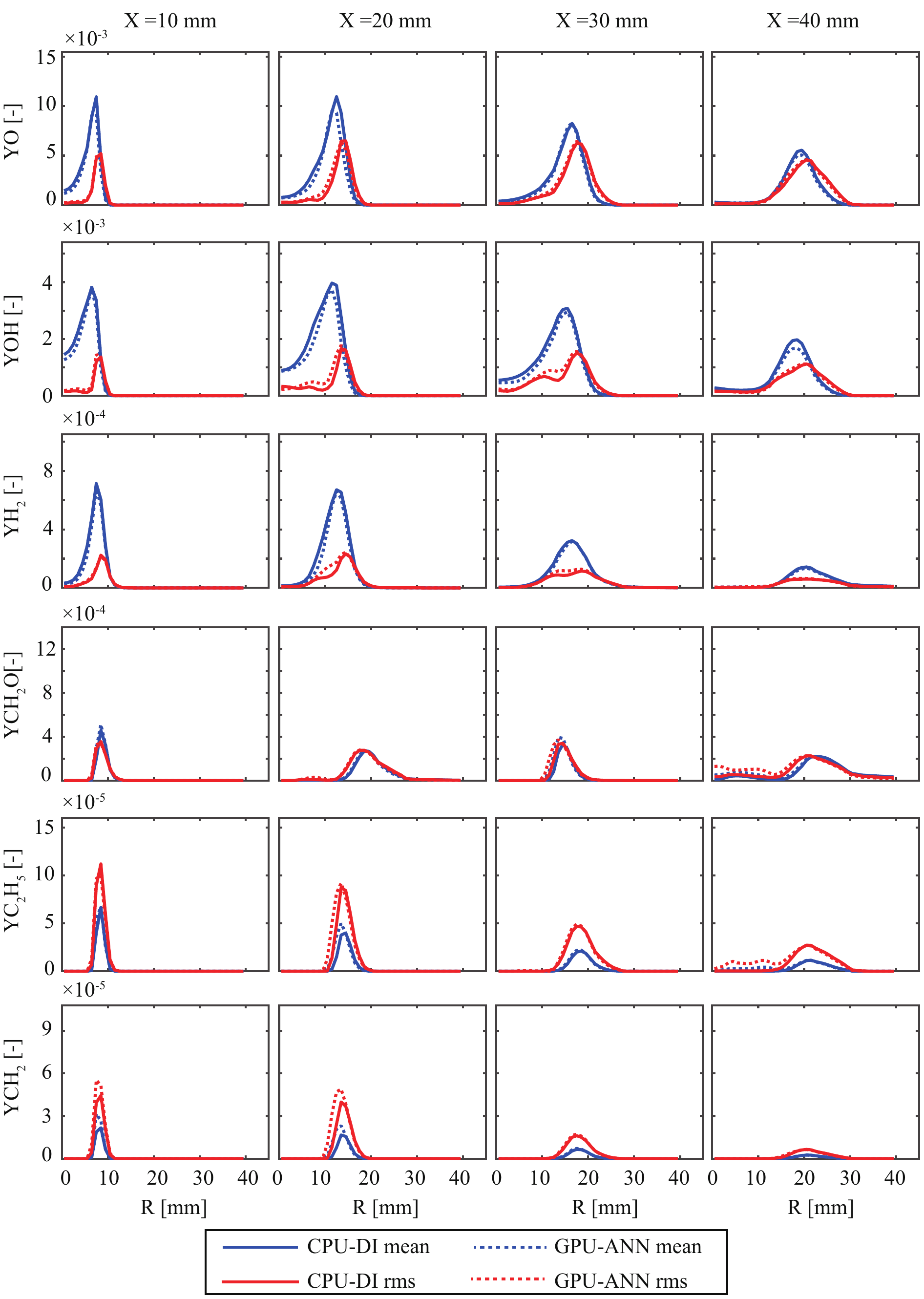}
\caption{\label{fig:11}Radial profiles of mean and RMS for minor species mass fraction across the order from $10^{-3}$ to $10^{-5}$ at four different axial locations in the experiment, CPU-DI, and GPU-ANN cases.}
\end{figure*}

While Section~\ref{sec3} has already shown that the macro flame structure can be accurately captured with the GPU-ANN approach in the swirling flame, it is unclear whether the statistical characteristics can be accurately captured in the more complicated swirling stratified flame case~(SWB7). This will be explored in the subsequent discussion. The mean and RMS statistics for the SWB7 case are presented in Fig.~\ref{fig:10} and~\ref{fig:11}.  
The mean and RMS values are identified using the black and red colors, respectively. Solid lines illustrate outcomes from the CPU-DI approach, and dashed lines correspond to the GPU-ANN approach. The circles are used to denote the experimental mean and RMS data~\cite{sweeney2012struc}. Fig.~\ref{fig:10} shows the simulated scalar profiles of the temperature, mass fraction of \ce{CH4}, carbon dioxide~(\ce{CO2}), oxygen~(\ce{O2}), Carbon monoxide~(\ce{CO}), and water~(\ce{H2O}) and compares them to the corresponding experimental data at four different axial locations. It is clear that the simulated results from the GPU-ANN approach are almost identical to those from the CPU-DI approach. This suggests that the GPU-ANN method is capable of replicating the key combustion characteristics modeled by the traditional CPU-DI approach even under the complicated swirling flow conditions. As compared to the experimental results, good agreements between the experimental and simulated results are observed from axial location $10\,\mathrm{mm}$ to $30\,\mathrm{mm}$. However, there some discrepancies are shown for $40\,\mathrm{mm}$. Specifically, the peak value of the temperature is under-predicted by $200\,\mathrm{K}$. This consequently leads to a lower prediction of \ce{H2O} and \ce{CO2}. The possible reason could be attributed to an overestimation of the recirculation flow intensity, leading to enhanced mixing with the air downstream~\cite{turkeri2021}. This is evident from the observed profiles of \ce{O2}, where a higher concentration of \ce{O2} is predicted within the recirculation zone. 
It is important to mention that the primary goal of this work is not to directly compare simulation outcomes with the experimental data. A further investigation of the effects of swirl on the flow and flame structure of the recirculation zone falls beyond the scope of this study.

\begin{figure*}
\includegraphics[width=430pt]{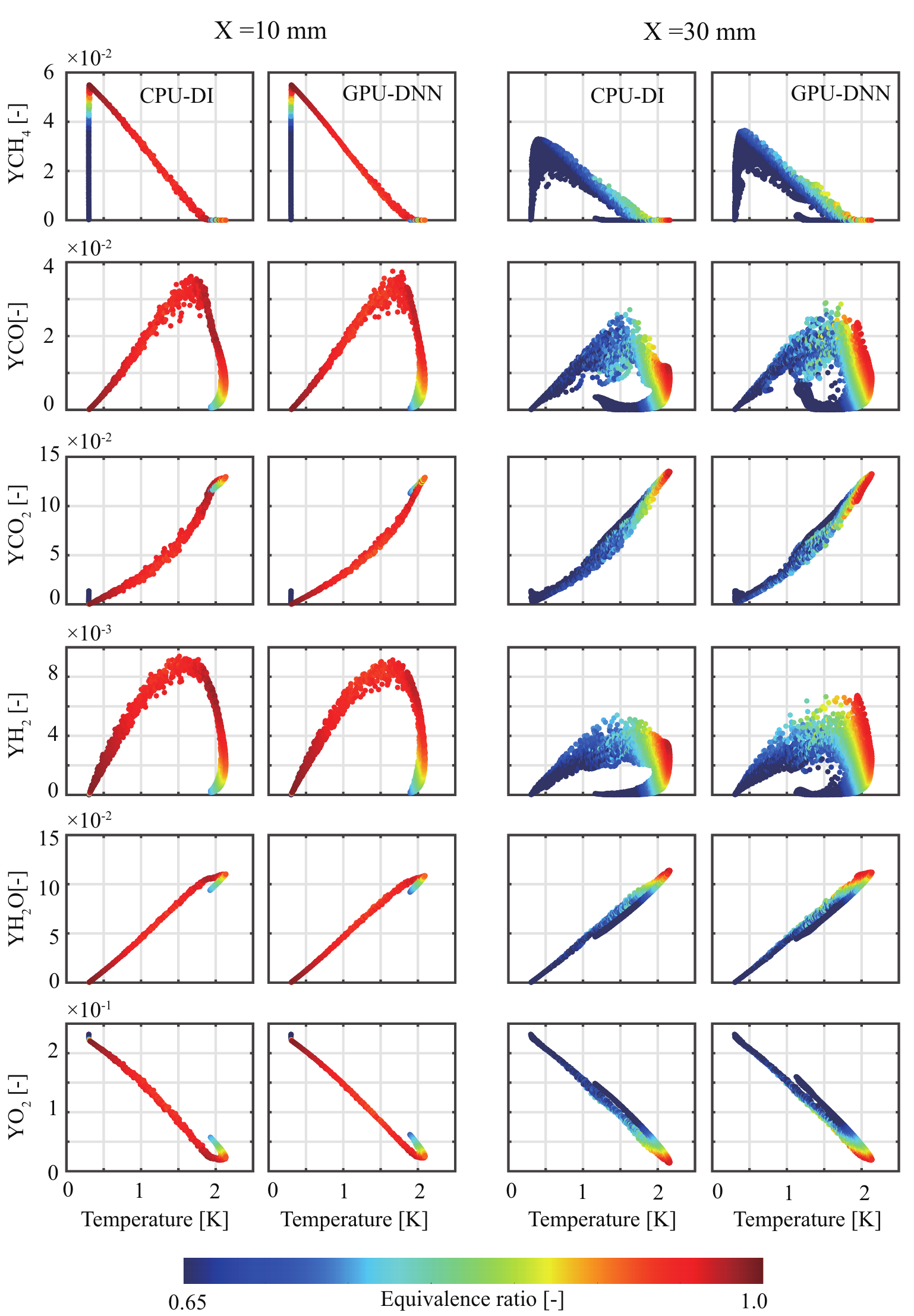}
\caption{\label{fig:12}Instantaneous scatter plots of key species mass fraction over flame temperature for CPU-DI and GPU-ANN at axial locations $\text{X} = 10\,\mathrm{mm}, 30\,\mathrm{mm}$.}
\end{figure*}

The minor species in the chemical reaction system cover a wider range of orders of magnitude in the thermochemical phase space, presenting a more significant challenge to the ANN's predictive capabilities than the major species. Consequently, it is crucial to conduct a more thorough assessment of the GPU-ANN approach's accuracy in predicting these minor species. Fig.~\ref{fig:11} shows the radial mean and RMS profiles of several minor species with mass fractions spanning from the order of $10^{-3}$ to $10^{-5}$. Overall, the mass fraction of all species simulated by the GPU-ANN approach shows a good agreement with the simulation results obtained from the CPU-DI approach. However, it is clear that the predictive performance of the GPU-ANN approach for minor species does not match the strength demonstrated for major species, as illustrated in Fig.~\ref{fig:10}. Specifically, there are slight discrepancies in the descending order of the mass fractions. For instance, the mass fraction of methylene~(\ce{CH2}) with the order of $10^{-5}$ is under-predicted by approximately 10\,\% at the axial location of $10\,\mathrm{mm}$ and $20\,\mathrm{mm}$. 
In the study by Readshaw et al.~\cite{readshaw2023}, the predicted outcomes for minor species, whose mass fractions are on the order of $10^{-5}$ obtained using the ANN method, closely align with those derived from the DI method. This congruence is likely attributable to the RMSE of the ANN ($10^{-8}$) obtained in their study, compared to the order of $10^{-7}$ noted in the current study.
However, if there is no particular interest in these relatively small mass fractions of the species involved, the prediction error observed with the GPU-ANN approach is considered acceptable.

Beyond the statistical characteristics of the scalar, scatter distribution in temperature-species at a single instant is also a crucial metric for evaluating the performance of the GPU-ANN approach. Fig.~\ref{fig:12} compares the scatter plots for the mass fraction of \ce{CH4}, \ce{CO}, \ce{CO2}, \ce{H2}, \ce{H2O}, and \ce{O2} in the temperature space at axial location of $10\,\mathrm{mm}$ and $30\,\mathrm{mm}$ between the GPU-ANN and CPU-DI approaches.  
All scatters are colored by the equivalence ratio ranging from 0.65 to 1.0. At the axial location of $10\,\mathrm{mm}$, most of the scatter points are concentrated around the stoichiometric state, with only a few scatters around the equivalence ratio of 0.8 for both the GPU-ANN and CPU-DI approaches. 
As the downstream distance increases to $30\,\mathrm{mm}$, numerous scatter points with low equivalence ratios are predicted by both GPU-ANN and CPU-DI approaches, particularly at temperatures greater than $1500\,\mathrm{K}$. This pattern, arising from entrainment, highlights the GPU-ANN approach's adeptness in capturing the local structure of the swirling flame.

\section{Computational saving}
\label{sec5}

\begin{figure}
\includegraphics[width=260pt]{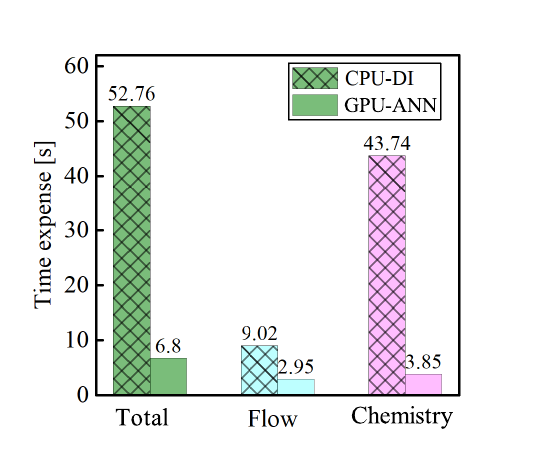}
\caption{\label{fig:13} Time expense for CPU-DI (32 cores) and GPU-ANN (1 card) approaches in the XJTU swirling flame case.}
\end{figure}

\begin{figure}
\includegraphics[width=250pt]{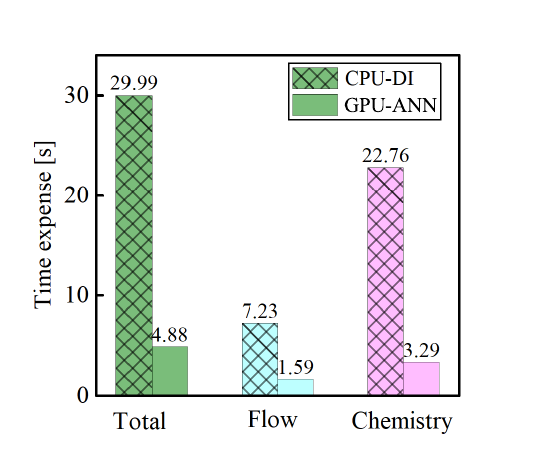}
\caption{\label{fig:14} Time expense for CPU-DI (32 cores) and GPU-ANN (1 card) approaches in the Cambridge swirling stratified flame case.}
\end{figure}

It is expected that the GPU-ANN approach is able to significantly decrease the computational cost. This is illustrated in Fig.~\ref{fig:13} and~\ref{fig:14}, where the total, flow, and chemistry time expenses of a single time step are compared between the CPU-DI and GPU-ANN approaches for the XJTU and SWB7 swirling flame cases. 
The cases involving the GPU-ANN are executed using a single GPU card, whereas the cases employing the CPU-DI method are conducted on a node with 32 CPU cores. 
It should be noted that training the ANN for 1600 epochs on a single GPU card took less than 2 hours, which is negligible compared to the overall computational costs. The training cost is hence not included in Fig.~\ref{fig:13} and~\ref{fig:14}.
Since the ANN is employed solely to replace the integration of ODEs, the computational savings in chemistry calculations are primarily attributed to the ANN, whereas savings in flow computations are mainly due to the GPU.
As shown in Fig.~\ref{fig:13} and~\ref{fig:14}, the use of GPU-ANN significantly enhances computational efficiency for both the XJTU and Cambridge swirling flames. With the GPU-ANN, the XJTU swirling flame sees a time reduction of 91.2\,\% for chemical reactions and 67.3\,\% for flow computations, giving a 1GPU-to-1CPU speed-up ratio of 363 and 98 compared to the CPU-DI method. Meanwhile, the Cambridge swirling stratified flame benefits from 85.5\,\% and 78.0\,\% time savings in chemistry and flow calculations, respectively, with speed-ups of 221:1 and 145:1 under the same comparison.
In these two cases of swirling flames, the computation of a single step with one GPU-ANN card is approximately seven times quicker than using a CPU-DI with 32 cores. 
Although there is a marked reduction in time spent on chemistry sources, it still accounts for 60\,\% of the total time in both swirling flame cases. This proportion may rise with the scaling of inference points. Consequently, optimizing the inference framework and minimizing the ANN's parameters is essential to achieve further speedup in future work.

\section{Conclusion}
\label{sec6}
In the present study, two LES of swirling flames including the XJTU swirling flame and the Cambridge swirling stratified flame are conducted to evaluate the proposed GPU-ANN approach. This method involves the ANN model that is trained using 1-D laminar flame solutions augmented with random perturbations and then implemented within the solver that is fully accelerated by GPU and based on OpenFOAM. 

The results of the simulations reveal a strong alignment between the GPU-ANN method and the traditional approach using the CPU solver with direct integration of chemistry, in terms of capturing the macro flame structure and statistical characteristics such as average and RMS of scalar fields and scattered instantaneous points. This consistency suggests that the GPU-ANN method can achieve a level of accuracy comparable to that of the standard CPU-DI solver even in complex scenarios. 
Furthermore, the computational savings achieved by the GPU-ANN approach have been evaluated. The results demonstrate that for the Cambridge swirling flame, time spent on chemistry and flow calculations is reduced by 91.2\,\% and 67.3\,\%, respectively, when using one GPU card as opposed to 32 CPU cores. For the XJTU swirling flame, the corresponding time reductions of 85.5\,\% for chemistry and 78.0\,\% for flow computations were observed under the same comparison.
This corresponds to a speed-up factor exceeding two orders of magnitude.

The present study showcases the potential for widespread application of the proposed GPU-ANN approach in combustion simulations that tackle various complex scenarios using detailed chemistry, while also substantially cutting computational costs. However, it should be noted that the time spent on chemistry sources still accounts for 60\,\% of the total time. Future work will be focused on optimizing the inference framework and minimizing the ANN’s parameters. 

\begin{acknowledgments}
This work is supported by the National Natural Science Foundation of China (Grant Nos. 92270203 and 52276096), GHfund C (202302032372), CCF-Baidu Open Fund, the “Emerging Engineering Interdisciplinary-Young Scholars Project, Peking University", and the Fundamental Research ARCHER2 UK National Supercomputing Service (https://www.archer2.ac.uk). Part of the numerical simulations was performed on the High Funds for the Central Universities. The support from the Royal Society (IEC/NSFC/223421) is also acknowledged. This work used the Performance Computing Platform of CAPT of Peking University.
\end{acknowledgments}

\section*{Author declarations}
\subsection*{Conflict of Interest}
The authors have no conflicts to disclose.
\subsection*{Author Contributions}
\textbf{Min Zhang:} Conceptualization (lead), Software (equal), Validation (lead), Writing - original draft (lead); \textbf{Runze Mao:} Software (equal), Conceptualization (equal); \textbf{Han Li:} Methodology (equal), Writing - original draft (supporting); \textbf{Zhenhua An:} Resource (equal), Software (equal); \textbf{Zhi X. Chen:} Funding acquisition (lead), Supervision (lead), Software (equal), Writing - review \& editing (lead);

\section*{Data Availability Statement}

The underlying data supporting the conclusions of this research can be obtained from the corresponding author if requested with justified cause.

\appendix

\bibliography{aipsamp}

\providecommand{\noopsort}[1]{}\providecommand{\singleletter}[1]{#1}%
\begin{thebibliography}{56}%
\makeatletter
\providecommand \@ifxundefined [1]{%
 \@ifx{#1\undefined}
}%
\providecommand \@ifnum [1]{%
 \ifnum #1\expandafter \@firstoftwo
 \else \expandafter \@secondoftwo
 \fi
}%
\providecommand \@ifx [1]{%
 \ifx #1\expandafter \@firstoftwo
 \else \expandafter \@secondoftwo
 \fi
}%
\providecommand \natexlab [1]{#1}%
\providecommand \enquote  [1]{``#1''}%
\providecommand \bibnamefont  [1]{#1}%
\providecommand \bibfnamefont [1]{#1}%
\providecommand \citenamefont [1]{#1}%
\providecommand \href@noop [0]{\@secondoftwo}%
\providecommand \href [0]{\begingroup \@sanitize@url \@href}%
\providecommand \@href[1]{\@@startlink{#1}\@@href}%
\providecommand \@@href[1]{\endgroup#1\@@endlink}%
\providecommand \@sanitize@url [0]{\catcode `\\12\catcode `\$12\catcode `\&12\catcode `\#12\catcode `\^12\catcode `\_12\catcode `\%12\relax}%
\providecommand \@@startlink[1]{}%
\providecommand \@@endlink[0]{}%
\providecommand \url  [0]{\begingroup\@sanitize@url \@url }%
\providecommand \@url [1]{\endgroup\@href {#1}{\urlprefix }}%
\providecommand \urlprefix  [0]{URL }%
\providecommand \Eprint [0]{\href }%
\providecommand \doibase [0]{http://dx.doi.org/}%
\providecommand \selectlanguage [0]{\@gobble}%
\providecommand \bibinfo  [0]{\@secondoftwo}%
\providecommand \bibfield  [0]{\@secondoftwo}%
\providecommand \translation [1]{[#1]}%
\providecommand \BibitemOpen [0]{}%
\providecommand \bibitemStop [0]{}%
\providecommand \bibitemNoStop [0]{.\EOS\space}%
\providecommand \EOS [0]{\spacefactor3000\relax}%
\providecommand \BibitemShut  [1]{\csname bibitem#1\endcsname}%
\let\auto@bib@innerbib\@empty
\bibitem [{\citenamefont {Avdi{\'c}}\ \emph {et~al.}(2017)\citenamefont {Avdi{\'c}}, \citenamefont {Kuenne}, \citenamefont {di~Mare},\ and\ \citenamefont {Janicka}}]{avdic2017}%
  \BibitemOpen
  \bibfield  {author} {\bibinfo {author} {\bibfnamefont {A.}~\bibnamefont {Avdi{\'c}}}, \bibinfo {author} {\bibfnamefont {G.}~\bibnamefont {Kuenne}}, \bibinfo {author} {\bibfnamefont {F.}~\bibnamefont {di~Mare}}, \ and\ \bibinfo {author} {\bibfnamefont {J.}~\bibnamefont {Janicka}},\ }\bibfield  {title} {\enquote {\bibinfo {title} {{LES} combustion modeling using the eulerian stochastic field method coupled with tabulated chemistry},}\ }\href@noop {} {\bibfield  {journal} {\bibinfo  {journal} {Combust. Flame}\ }\textbf {\bibinfo {volume} {175}},\ \bibinfo {pages} {201--219} (\bibinfo {year} {2017})}\BibitemShut {NoStop}%
\bibitem [{\citenamefont {Peters}(2001)}]{Peters2001}%
  \BibitemOpen
  \bibfield  {author} {\bibinfo {author} {\bibfnamefont {N.}~\bibnamefont {Peters}},\ }\href@noop {} {\emph {\bibinfo {title} {Turbulent combustion}}}\ (\bibinfo  {publisher} {Measurement Science and Technology},\ \bibinfo {year} {2001})\BibitemShut {NoStop}%
\bibitem [{\citenamefont {Chen}, \citenamefont {Kollmann},\ and\ \citenamefont {Dibble}(1989)}]{chen1989}%
  \BibitemOpen
  \bibfield  {author} {\bibinfo {author} {\bibfnamefont {J.}~\bibnamefont {Chen}}, \bibinfo {author} {\bibfnamefont {W.}~\bibnamefont {Kollmann}}, \ and\ \bibinfo {author} {\bibfnamefont {R.}~\bibnamefont {Dibble}},\ }\bibfield  {title} {\enquote {\bibinfo {title} {{PDF} modeling of turbulent nonpremixed methane jet flames},}\ }\href@noop {} {\bibfield  {journal} {\bibinfo  {journal} {Combust. Sci. Technol.}\ }\textbf {\bibinfo {volume} {64}},\ \bibinfo {pages} {315--346} (\bibinfo {year} {1989})}\BibitemShut {NoStop}%
\bibitem [{\citenamefont {Van~Oijen}\ and\ \citenamefont {De~Goey}(2000)}]{van2000}%
  \BibitemOpen
  \bibfield  {author} {\bibinfo {author} {\bibfnamefont {J.}~\bibnamefont {Van~Oijen}}\ and\ \bibinfo {author} {\bibfnamefont {L.}~\bibnamefont {De~Goey}},\ }\bibfield  {title} {\enquote {\bibinfo {title} {Modelling of premixed laminar flames using flamelet-generated manifolds},}\ }\href@noop {} {\bibfield  {journal} {\bibinfo  {journal} {Combust. Sci. Technol.}\ }\textbf {\bibinfo {volume} {161}},\ \bibinfo {pages} {113--137} (\bibinfo {year} {2000})}\BibitemShut {NoStop}%
\bibitem [{\citenamefont {Zhang}\ \emph {et~al.}(2021{\natexlab{a}})\citenamefont {Zhang}, \citenamefont {Karaca}, \citenamefont {Wang}, \citenamefont {Huang},\ and\ \citenamefont {van Oijen}}]{zhang2021}%
  \BibitemOpen
  \bibfield  {author} {\bibinfo {author} {\bibfnamefont {W.}~\bibnamefont {Zhang}}, \bibinfo {author} {\bibfnamefont {S.}~\bibnamefont {Karaca}}, \bibinfo {author} {\bibfnamefont {J.}~\bibnamefont {Wang}}, \bibinfo {author} {\bibfnamefont {Z.}~\bibnamefont {Huang}}, \ and\ \bibinfo {author} {\bibfnamefont {J.}~\bibnamefont {van Oijen}},\ }\bibfield  {title} {\enquote {\bibinfo {title} {Large eddy simulation of the cambridge/sandia stratified flame with flamelet-generated manifolds: Effects of non-unity lewis numbers and stretch},}\ }\href@noop {} {\bibfield  {journal} {\bibinfo  {journal} {Combust. Flame}\ }\textbf {\bibinfo {volume} {227}},\ \bibinfo {pages} {106--119} (\bibinfo {year} {2021}{\natexlab{a}})}\BibitemShut {NoStop}%
\bibitem [{\citenamefont {Baik}\ \emph {et~al.}(2022)\citenamefont {Baik}, \citenamefont {Inanc}, \citenamefont {Rieth},\ and\ \citenamefont {Kempf}}]{baik2022}%
  \BibitemOpen
  \bibfield  {author} {\bibinfo {author} {\bibfnamefont {S.-J.}\ \bibnamefont {Baik}}, \bibinfo {author} {\bibfnamefont {E.}~\bibnamefont {Inanc}}, \bibinfo {author} {\bibfnamefont {M.}~\bibnamefont {Rieth}}, \ and\ \bibinfo {author} {\bibfnamefont {A.}~\bibnamefont {Kempf}},\ }\bibfield  {title} {\enquote {\bibinfo {title} {Lagrangian filtered density function modeling of a turbulent stratified flame combined with flamelet approach},}\ }\href@noop {} {\bibfield  {journal} {\bibinfo  {journal} {Phys. Fluids}\ }\textbf {\bibinfo {volume} {34}} (\bibinfo {year} {2022})}\BibitemShut {NoStop}%
\bibitem [{\citenamefont {Maas}\ and\ \citenamefont {Pope}(1992)}]{Maas1992}%
  \BibitemOpen
  \bibfield  {author} {\bibinfo {author} {\bibfnamefont {U.}~\bibnamefont {Maas}}\ and\ \bibinfo {author} {\bibfnamefont {S.~B.}\ \bibnamefont {Pope}},\ }\bibfield  {title} {\enquote {\bibinfo {title} {Implementation of simplified chemical kinetics based on intrinsic low-dimensional manifolds},}\ }\href@noop {} {\bibfield  {journal} {\bibinfo  {journal} {Symposium (International) on Combustion}\ }\textbf {\bibinfo {volume} {24}},\ \bibinfo {pages} {103--112} (\bibinfo {year} {1992})}\BibitemShut {NoStop}%
\bibitem [{\citenamefont {Pope}(1997)}]{pope1997}%
  \BibitemOpen
  \bibfield  {author} {\bibinfo {author} {\bibfnamefont {S.~B.}\ \bibnamefont {Pope}},\ }\bibfield  {title} {\enquote {\bibinfo {title} {Computationally efficient implementation of combustion chemistry using in situ adaptive tabulation},}\ }\href@noop {} {\bibfield  {journal} {\bibinfo  {journal} {Combust. Theory Model}\ }\textbf {\bibinfo {volume} {1}},\ \bibinfo {pages} {41--63} (\bibinfo {year} {1997})}\BibitemShut {NoStop}%
\bibitem [{\citenamefont {Ketelheun}\ \emph {et~al.}(2011)\citenamefont {Ketelheun}, \citenamefont {Olbricht}, \citenamefont {Hahn},\ and\ \citenamefont {Janicka}}]{ketelheun2011}%
  \BibitemOpen
  \bibfield  {author} {\bibinfo {author} {\bibfnamefont {A.}~\bibnamefont {Ketelheun}}, \bibinfo {author} {\bibfnamefont {C.}~\bibnamefont {Olbricht}}, \bibinfo {author} {\bibfnamefont {F.}~\bibnamefont {Hahn}}, \ and\ \bibinfo {author} {\bibfnamefont {J.}~\bibnamefont {Janicka}},\ }\bibfield  {title} {\enquote {\bibinfo {title} {No prediction in turbulent flames using les/fgm with additional transport equations},}\ }\href@noop {} {\bibfield  {journal} {\bibinfo  {journal} {Prog. Energy Combust. Sci.}\ }\textbf {\bibinfo {volume} {3}},\ \bibinfo {pages} {2975--2982} (\bibinfo {year} {2011})}\BibitemShut {NoStop}%
\bibitem [{\citenamefont {Zhou}\ \emph {et~al.}(2022)\citenamefont {Zhou}, \citenamefont {Song}, \citenamefont {Ji},\ and\ \citenamefont {Wei}}]{Zhou2022}%
  \BibitemOpen
  \bibfield  {author} {\bibinfo {author} {\bibfnamefont {L.}~\bibnamefont {Zhou}}, \bibinfo {author} {\bibfnamefont {Y.}~\bibnamefont {Song}}, \bibinfo {author} {\bibfnamefont {W.}~\bibnamefont {Ji}}, \ and\ \bibinfo {author} {\bibfnamefont {H.}~\bibnamefont {Wei}},\ }\bibfield  {title} {\enquote {\bibinfo {title} {Machine learning for combustion},}\ }\href@noop {} {\bibfield  {journal} {\bibinfo  {journal} {Energy and AI}\ }\textbf {\bibinfo {volume} {7}},\ \bibinfo {pages} {100128} (\bibinfo {year} {2022})}\BibitemShut {NoStop}%
\bibitem [{\citenamefont {Ihme}, \citenamefont {Chung},\ and\ \citenamefont {Mishra}(2022)}]{Ihme2022}%
  \BibitemOpen
  \bibfield  {author} {\bibinfo {author} {\bibfnamefont {M.}~\bibnamefont {Ihme}}, \bibinfo {author} {\bibfnamefont {W.~T.}\ \bibnamefont {Chung}}, \ and\ \bibinfo {author} {\bibfnamefont {A.~A.}\ \bibnamefont {Mishra}},\ }\bibfield  {title} {\enquote {\bibinfo {title} {Combustion machine learning: Principles, progress and prospects},}\ }\href@noop {} {\bibfield  {journal} {\bibinfo  {journal} {Prog. Energy Combust. Sci.}\ }\textbf {\bibinfo {volume} {91}},\ \bibinfo {pages} {101010} (\bibinfo {year} {2022})}\BibitemShut {NoStop}%
\bibitem [{\citenamefont {Christo}\ \emph {et~al.}(1995)\citenamefont {Christo}, \citenamefont {Masri}, \citenamefont {Nebot},\ and\ \citenamefont {Tur{\'a}nyi}}]{christo1995}%
  \BibitemOpen
  \bibfield  {author} {\bibinfo {author} {\bibfnamefont {F.}~\bibnamefont {Christo}}, \bibinfo {author} {\bibfnamefont {A.}~\bibnamefont {Masri}}, \bibinfo {author} {\bibfnamefont {E.}~\bibnamefont {Nebot}}, \ and\ \bibinfo {author} {\bibfnamefont {T.}~\bibnamefont {Tur{\'a}nyi}},\ }\bibfield  {title} {\enquote {\bibinfo {title} {Utilising artificial neural network and repro-modelling in turbulent combustion},}\ }\href@noop {} {\bibfield  {journal} {\bibinfo  {journal} {Proceedings of ICNN'95-International Conference on Neural Networks}\ }\textbf {\bibinfo {volume} {2}},\ \bibinfo {pages} {911--916} (\bibinfo {year} {1995})}\BibitemShut {NoStop}%
\bibitem [{\citenamefont {Brown}\ \emph {et~al.}(2021)\citenamefont {Brown}, \citenamefont {Antil}, \citenamefont {L{\"o}hner}, \citenamefont {Togashi},\ and\ \citenamefont {Verma}}]{brown2021}%
  \BibitemOpen
  \bibfield  {author} {\bibinfo {author} {\bibfnamefont {T.~S.}\ \bibnamefont {Brown}}, \bibinfo {author} {\bibfnamefont {H.}~\bibnamefont {Antil}}, \bibinfo {author} {\bibfnamefont {R.}~\bibnamefont {L{\"o}hner}}, \bibinfo {author} {\bibfnamefont {F.}~\bibnamefont {Togashi}}, \ and\ \bibinfo {author} {\bibfnamefont {D.}~\bibnamefont {Verma}},\ }\bibfield  {title} {\enquote {\bibinfo {title} {Novel {DNN}s for stiff {ODE}s with applications to chemically reacting flows},}\ }\href@noop {} {\bibfield  {journal} {\bibinfo  {journal} {High Performance Computing: ISC High Performance Digital 2021 International Workshops}\ ,\ \bibinfo {pages} {23--39}} (\bibinfo {year} {2021})}\BibitemShut {NoStop}%
\bibitem [{\citenamefont {Petersen}\ and\ \citenamefont {Hanson}(1999)}]{petersen1999}%
  \BibitemOpen
  \bibfield  {author} {\bibinfo {author} {\bibfnamefont {E.~L.}\ \bibnamefont {Petersen}}\ and\ \bibinfo {author} {\bibfnamefont {R.~K.}\ \bibnamefont {Hanson}},\ }\bibfield  {title} {\enquote {\bibinfo {title} {Reduced kinetics mechanisms for ram accelerator combustion},}\ }\href@noop {} {\bibfield  {journal} {\bibinfo  {journal} {J. Propuls. Power}\ }\textbf {\bibinfo {volume} {15}},\ \bibinfo {pages} {591--600} (\bibinfo {year} {1999})}\BibitemShut {NoStop}%
\bibitem [{\citenamefont {Blasco}\ \emph {et~al.}(2000)\citenamefont {Blasco}, \citenamefont {Fueyo}, \citenamefont {Dopazo},\ and\ \citenamefont {Chen}}]{blasco2000}%
  \BibitemOpen
  \bibfield  {author} {\bibinfo {author} {\bibfnamefont {J.~A.}\ \bibnamefont {Blasco}}, \bibinfo {author} {\bibfnamefont {N.}~\bibnamefont {Fueyo}}, \bibinfo {author} {\bibfnamefont {C.}~\bibnamefont {Dopazo}}, \ and\ \bibinfo {author} {\bibfnamefont {J.}~\bibnamefont {Chen}},\ }\bibfield  {title} {\enquote {\bibinfo {title} {A self-organizing-map approach to chemistry representation in combustion applications},}\ }\href@noop {} {\bibfield  {journal} {\bibinfo  {journal} {Combust. Theory Model.}\ }\textbf {\bibinfo {volume} {4}},\ \bibinfo {pages} {61} (\bibinfo {year} {2000})}\BibitemShut {NoStop}%
\bibitem [{\citenamefont {Blasco}\ \emph {et~al.}(1998)\citenamefont {Blasco}, \citenamefont {Fueyo}, \citenamefont {Dopazo},\ and\ \citenamefont {Ballester}}]{blasco1998}%
  \BibitemOpen
  \bibfield  {author} {\bibinfo {author} {\bibfnamefont {J.}~\bibnamefont {Blasco}}, \bibinfo {author} {\bibfnamefont {N.}~\bibnamefont {Fueyo}}, \bibinfo {author} {\bibfnamefont {C.}~\bibnamefont {Dopazo}}, \ and\ \bibinfo {author} {\bibfnamefont {J.}~\bibnamefont {Ballester}},\ }\bibfield  {title} {\enquote {\bibinfo {title} {Modelling the temporal evolution of a reduced combustion chemical system with an artificial neural network},}\ }\href@noop {} {\bibfield  {journal} {\bibinfo  {journal} {Combust. Flame}\ }\textbf {\bibinfo {volume} {113}},\ \bibinfo {pages} {38--52} (\bibinfo {year} {1998})}\BibitemShut {NoStop}%
\bibitem [{\citenamefont {Parente}\ \emph {et~al.}(2009)\citenamefont {Parente}, \citenamefont {Sutherland}, \citenamefont {Tognotti},\ and\ \citenamefont {Smith}}]{parente2009}%
  \BibitemOpen
  \bibfield  {author} {\bibinfo {author} {\bibfnamefont {A.}~\bibnamefont {Parente}}, \bibinfo {author} {\bibfnamefont {J.~C.}\ \bibnamefont {Sutherland}}, \bibinfo {author} {\bibfnamefont {L.}~\bibnamefont {Tognotti}}, \ and\ \bibinfo {author} {\bibfnamefont {P.~J.}\ \bibnamefont {Smith}},\ }\bibfield  {title} {\enquote {\bibinfo {title} {Identification of low-dimensional manifolds in turbulent flames},}\ }\href@noop {} {\bibfield  {journal} {\bibinfo  {journal} {Prog. Energy Combust. Sci}\ }\textbf {\bibinfo {volume} {32}},\ \bibinfo {pages} {1579--1586} (\bibinfo {year} {2009})}\BibitemShut {NoStop}%
\bibitem [{\citenamefont {Abdelwahid}\ \emph {et~al.}(2023)\citenamefont {Abdelwahid}, \citenamefont {Malik}, \citenamefont {Hammoud}, \citenamefont {Hern{\'a}ndez-P{\'e}rez}, \citenamefont {Ghanem},\ and\ \citenamefont {Im}}]{abdelwahid2023}%
  \BibitemOpen
  \bibfield  {author} {\bibinfo {author} {\bibfnamefont {S.}~\bibnamefont {Abdelwahid}}, \bibinfo {author} {\bibfnamefont {M.~R.}\ \bibnamefont {Malik}}, \bibinfo {author} {\bibfnamefont {H.~A. A.~K.}\ \bibnamefont {Hammoud}}, \bibinfo {author} {\bibfnamefont {F.~E.}\ \bibnamefont {Hern{\'a}ndez-P{\'e}rez}}, \bibinfo {author} {\bibfnamefont {B.}~\bibnamefont {Ghanem}}, \ and\ \bibinfo {author} {\bibfnamefont {H.~G.}\ \bibnamefont {Im}},\ }\bibfield  {title} {\enquote {\bibinfo {title} {Large eddy simulations of ammonia-hydrogen jet flames at elevated pressure using principal component analysis and deep neural networks},}\ }\href@noop {} {\bibfield  {journal} {\bibinfo  {journal} {Combust. Flame}\ }\textbf {\bibinfo {volume} {253}},\ \bibinfo {pages} {112781} (\bibinfo {year} {2023})}\BibitemShut {NoStop}%
\bibitem [{\citenamefont {Castellanos}\ \emph {et~al.}(2023)\citenamefont {Castellanos}, \citenamefont {SM~Freitas}, \citenamefont {Parente},\ and\ \citenamefont {Contino}}]{castellanos2023}%
  \BibitemOpen
  \bibfield  {author} {\bibinfo {author} {\bibfnamefont {L.}~\bibnamefont {Castellanos}}, \bibinfo {author} {\bibfnamefont {R.}~\bibnamefont {SM~Freitas}}, \bibinfo {author} {\bibfnamefont {A.}~\bibnamefont {Parente}}, \ and\ \bibinfo {author} {\bibfnamefont {F.}~\bibnamefont {Contino}},\ }\bibfield  {title} {\enquote {\bibinfo {title} {Deep learning dynamical latencies for the analysis and reduction of combustion chemistry kinetics},}\ }\href@noop {} {\bibfield  {journal} {\bibinfo  {journal} {Phys. Fluids}\ }\textbf {\bibinfo {volume} {35}} (\bibinfo {year} {2023})}\BibitemShut {NoStop}%
\bibitem [{\citenamefont {Goswami}\ \emph {et~al.}(2024)\citenamefont {Goswami}, \citenamefont {Jagtap}, \citenamefont {Babaee}, \citenamefont {Susi},\ and\ \citenamefont {Karniadakis}}]{Goswami2024}%
  \BibitemOpen
  \bibfield  {author} {\bibinfo {author} {\bibfnamefont {S.}~\bibnamefont {Goswami}}, \bibinfo {author} {\bibfnamefont {A.~D.}\ \bibnamefont {Jagtap}}, \bibinfo {author} {\bibfnamefont {H.}~\bibnamefont {Babaee}}, \bibinfo {author} {\bibfnamefont {B.~T.}\ \bibnamefont {Susi}}, \ and\ \bibinfo {author} {\bibfnamefont {G.~E.}\ \bibnamefont {Karniadakis}},\ }\bibfield  {title} {\enquote {\bibinfo {title} {Learning stiff chemical kinetics using extended deep neural operators},}\ }\href@noop {} {\bibfield  {journal} {\bibinfo  {journal} {Comput. Methods Appl. Mech. Eng.}\ }\textbf {\bibinfo {volume} {419}},\ \bibinfo {pages} {116674} (\bibinfo {year} {2024})}\BibitemShut {NoStop}%
\bibitem [{\citenamefont {An}\ \emph {et~al.}(2020)\citenamefont {An}, \citenamefont {He}, \citenamefont {Luo}, \citenamefont {Qin},\ and\ \citenamefont {Liu}}]{an2020}%
  \BibitemOpen
  \bibfield  {author} {\bibinfo {author} {\bibfnamefont {J.}~\bibnamefont {An}}, \bibinfo {author} {\bibfnamefont {G.}~\bibnamefont {He}}, \bibinfo {author} {\bibfnamefont {K.}~\bibnamefont {Luo}}, \bibinfo {author} {\bibfnamefont {F.}~\bibnamefont {Qin}}, \ and\ \bibinfo {author} {\bibfnamefont {B.}~\bibnamefont {Liu}},\ }\bibfield  {title} {\enquote {\bibinfo {title} {Artificial neural network based chemical mechanisms for computationally efficient modeling of hydrogen/carbon monoxide/kerosene combustion},}\ }\href@noop {} {\bibfield  {journal} {\bibinfo  {journal} {Int. J. Hydrog. Energy}\ }\textbf {\bibinfo {volume} {45}},\ \bibinfo {pages} {29594--29605} (\bibinfo {year} {2020})}\BibitemShut {NoStop}%
\bibitem [{\citenamefont {Wenga}\ \emph {et~al.}(2023)\citenamefont {Wenga}, \citenamefont {Zhaoa}, \citenamefont {Zhanga},\ and\ \citenamefont {Zhoua}}]{wenga2023}%
  \BibitemOpen
  \bibfield  {author} {\bibinfo {author} {\bibfnamefont {Y.}~\bibnamefont {Wenga}}, \bibinfo {author} {\bibfnamefont {Z.}~\bibnamefont {Zhaoa}}, \bibinfo {author} {\bibfnamefont {H.}~\bibnamefont {Zhanga}}, \ and\ \bibinfo {author} {\bibfnamefont {D.}~\bibnamefont {Zhoua}},\ }\bibfield  {title} {\enquote {\bibinfo {title} {Extending fourier neural operators to learn stiff chemical kinetics},}\ }\href@noop {} {\bibfield  {journal} {\bibinfo  {journal} {DOI: 10.13140/RG.2.2.31763.66086}\ } (\bibinfo {year} {2023})}\BibitemShut {NoStop}%
\bibitem [{\citenamefont {Zhang}\ \emph {et~al.}(2022)\citenamefont {Zhang}, \citenamefont {Yi}, \citenamefont {Xu}, \citenamefont {Chen}, \citenamefont {Zhang}, \citenamefont {Weinan},\ and\ \citenamefont {Xu}}]{zhang2022}%
  \BibitemOpen
  \bibfield  {author} {\bibinfo {author} {\bibfnamefont {T.}~\bibnamefont {Zhang}}, \bibinfo {author} {\bibfnamefont {Y.}~\bibnamefont {Yi}}, \bibinfo {author} {\bibfnamefont {Y.}~\bibnamefont {Xu}}, \bibinfo {author} {\bibfnamefont {Z.~X.}\ \bibnamefont {Chen}}, \bibinfo {author} {\bibfnamefont {Y.}~\bibnamefont {Zhang}}, \bibinfo {author} {\bibfnamefont {E.}~\bibnamefont {Weinan}}, \ and\ \bibinfo {author} {\bibfnamefont {Z.-Q.~J.}\ \bibnamefont {Xu}},\ }\bibfield  {title} {\enquote {\bibinfo {title} {A multi-scale sampling method for accurate and robust deep neural network to predict combustion chemical kinetics},}\ }\href@noop {} {\bibfield  {journal} {\bibinfo  {journal} {Combust. Flame}\ }\textbf {\bibinfo {volume} {245}},\ \bibinfo {pages} {112319} (\bibinfo {year} {2022})}\BibitemShut {NoStop}%
\bibitem [{\citenamefont {Xu}\ \emph {et~al.}(2024)\citenamefont {Xu}, \citenamefont {Yao}, \citenamefont {Yi}, \citenamefont {Hang}, \citenamefont {Zhang}, \citenamefont {Zhang} \emph {et~al.}}]{xu2024}%
  \BibitemOpen
  \bibfield  {author} {\bibinfo {author} {\bibfnamefont {Z.-Q.~J.}\ \bibnamefont {Xu}}, \bibinfo {author} {\bibfnamefont {J.}~\bibnamefont {Yao}}, \bibinfo {author} {\bibfnamefont {Y.}~\bibnamefont {Yi}}, \bibinfo {author} {\bibfnamefont {L.}~\bibnamefont {Hang}}, \bibinfo {author} {\bibfnamefont {Y.}~\bibnamefont {Zhang}}, \bibinfo {author} {\bibfnamefont {T.}~\bibnamefont {Zhang}},  \emph {et~al.},\ }\bibfield  {title} {\enquote {\bibinfo {title} {Solving multiscale dynamical systems by deep learning},}\ }\href@noop {} {\bibfield  {journal} {\bibinfo  {journal} {arXiv preprint arXiv:2401.01220}\ } (\bibinfo {year} {2024})}\BibitemShut {NoStop}%
\bibitem [{\citenamefont {Readshaw}\ \emph {et~al.}(2023)\citenamefont {Readshaw}, \citenamefont {Franke}, \citenamefont {Jones},\ and\ \citenamefont {Rigopoulos}}]{readshaw2023}%
  \BibitemOpen
  \bibfield  {author} {\bibinfo {author} {\bibfnamefont {T.}~\bibnamefont {Readshaw}}, \bibinfo {author} {\bibfnamefont {L.~L.}\ \bibnamefont {Franke}}, \bibinfo {author} {\bibfnamefont {W.}~\bibnamefont {Jones}}, \ and\ \bibinfo {author} {\bibfnamefont {S.}~\bibnamefont {Rigopoulos}},\ }\bibfield  {title} {\enquote {\bibinfo {title} {Simulation of turbulent premixed flames with machine learning - tabulated thermochemistry},}\ }\href@noop {} {\bibfield  {journal} {\bibinfo  {journal} {Combust. Flame}\ }\textbf {\bibinfo {volume} {258}},\ \bibinfo {pages} {113058} (\bibinfo {year} {2023})}\BibitemShut {NoStop}%
\bibitem [{\citenamefont {Ding}\ \emph {et~al.}(2021)\citenamefont {Ding}, \citenamefont {Readshaw}, \citenamefont {Rigopoulos},\ and\ \citenamefont {Jones}}]{ding2021}%
  \BibitemOpen
  \bibfield  {author} {\bibinfo {author} {\bibfnamefont {T.}~\bibnamefont {Ding}}, \bibinfo {author} {\bibfnamefont {T.}~\bibnamefont {Readshaw}}, \bibinfo {author} {\bibfnamefont {S.}~\bibnamefont {Rigopoulos}}, \ and\ \bibinfo {author} {\bibfnamefont {W.}~\bibnamefont {Jones}},\ }\bibfield  {title} {\enquote {\bibinfo {title} {Simulation of turbulent premixed flames with machine learning-tabulated thermochemistry},}\ }\href@noop {} {\bibfield  {journal} {\bibinfo  {journal} {Combust. Flame}\ }\textbf {\bibinfo {volume} {231}},\ \bibinfo {pages} {111493} (\bibinfo {year} {2021})}\BibitemShut {NoStop}%
\bibitem [{\citenamefont {Ding}, \citenamefont {Rigopoulos},\ and\ \citenamefont {Jones}(2022)}]{ding2022}%
  \BibitemOpen
  \bibfield  {author} {\bibinfo {author} {\bibfnamefont {T.}~\bibnamefont {Ding}}, \bibinfo {author} {\bibfnamefont {S.}~\bibnamefont {Rigopoulos}}, \ and\ \bibinfo {author} {\bibfnamefont {W.}~\bibnamefont {Jones}},\ }\bibfield  {title} {\enquote {\bibinfo {title} {Machine learning tabulation of thermochemistry of fuel blends},}\ }\href@noop {} {\bibfield  {journal} {\bibinfo  {journal} {Appl. Energy Combust. Sci.}\ }\textbf {\bibinfo {volume} {12}},\ \bibinfo {pages} {100086} (\bibinfo {year} {2022})}\BibitemShut {NoStop}%
\bibitem [{\citenamefont {Liu}\ \emph {et~al.}(2024)\citenamefont {Liu}, \citenamefont {Ding}, \citenamefont {Liu}, \citenamefont {Rigopoulos},\ and\ \citenamefont {Luo}}]{liu2024}%
  \BibitemOpen
  \bibfield  {author} {\bibinfo {author} {\bibfnamefont {A.}~\bibnamefont {Liu}}, \bibinfo {author} {\bibfnamefont {T.}~\bibnamefont {Ding}}, \bibinfo {author} {\bibfnamefont {R.}~\bibnamefont {Liu}}, \bibinfo {author} {\bibfnamefont {S.}~\bibnamefont {Rigopoulos}}, \ and\ \bibinfo {author} {\bibfnamefont {K.}~\bibnamefont {Luo}},\ }\bibfield  {title} {\enquote {\bibinfo {title} {Machine learning tabulation of thermochemistry for turbulent dimethyl ether ({DME}) flames},}\ }\href@noop {} {\bibfield  {journal} {\bibinfo  {journal} {Fuel}\ }\textbf {\bibinfo {volume} {360}},\ \bibinfo {pages} {130338} (\bibinfo {year} {2024})}\BibitemShut {NoStop}%
\bibitem [{\citenamefont {P{\'e}rez}\ \emph {et~al.}(2018)\citenamefont {P{\'e}rez}, \citenamefont {Mukhadiyev}, \citenamefont {Xu}, \citenamefont {Sow}, \citenamefont {Lee}, \citenamefont {Sankaran},\ and\ \citenamefont {Im}}]{perez2018}%
  \BibitemOpen
  \bibfield  {author} {\bibinfo {author} {\bibfnamefont {F.~E.~H.}\ \bibnamefont {P{\'e}rez}}, \bibinfo {author} {\bibfnamefont {N.}~\bibnamefont {Mukhadiyev}}, \bibinfo {author} {\bibfnamefont {X.}~\bibnamefont {Xu}}, \bibinfo {author} {\bibfnamefont {A.}~\bibnamefont {Sow}}, \bibinfo {author} {\bibfnamefont {B.~J.}\ \bibnamefont {Lee}}, \bibinfo {author} {\bibfnamefont {R.}~\bibnamefont {Sankaran}}, \ and\ \bibinfo {author} {\bibfnamefont {H.~G.}\ \bibnamefont {Im}},\ }\bibfield  {title} {\enquote {\bibinfo {title} {Direct numerical simulations of reacting flows with detailed chemistry using many-core/{GPU} acceleration},}\ }\href@noop {} {\bibfield  {journal} {\bibinfo  {journal} {Computers \& Fluids}\ }\textbf {\bibinfo {volume} {173}},\ \bibinfo {pages} {73--79} (\bibinfo {year} {2018})}\BibitemShut {NoStop}%
\bibitem [{\citenamefont {Bielawski}\ \emph {et~al.}(2023)\citenamefont {Bielawski}, \citenamefont {Barwey}, \citenamefont {Prakash},\ and\ \citenamefont {Raman}}]{bielawski2023}%
  \BibitemOpen
  \bibfield  {author} {\bibinfo {author} {\bibfnamefont {R.}~\bibnamefont {Bielawski}}, \bibinfo {author} {\bibfnamefont {S.}~\bibnamefont {Barwey}}, \bibinfo {author} {\bibfnamefont {S.}~\bibnamefont {Prakash}}, \ and\ \bibinfo {author} {\bibfnamefont {V.}~\bibnamefont {Raman}},\ }\bibfield  {title} {\enquote {\bibinfo {title} {Highly-scalable {GPU}-accelerated compressible reacting flow solver for modeling high-speed flows},}\ }\href@noop {} {\bibfield  {journal} {\bibinfo  {journal} {Computers \& Fluids}\ ,\ \bibinfo {pages} {105972}} (\bibinfo {year} {2023})}\BibitemShut {NoStop}%
\bibitem [{\citenamefont {Esclapez}\ \emph {et~al.}(2023)\citenamefont {Esclapez}, \citenamefont {Day}, \citenamefont {Bell}, \citenamefont {Felden}, \citenamefont {Gilet}, \citenamefont {Grout}, \citenamefont {de~Frahan}, \citenamefont {Motheau}, \citenamefont {Nonaka}, \citenamefont {Owen} \emph {et~al.}}]{Esclapez2023}%
  \BibitemOpen
  \bibfield  {author} {\bibinfo {author} {\bibfnamefont {L.}~\bibnamefont {Esclapez}}, \bibinfo {author} {\bibfnamefont {M.}~\bibnamefont {Day}}, \bibinfo {author} {\bibfnamefont {J.}~\bibnamefont {Bell}}, \bibinfo {author} {\bibfnamefont {A.}~\bibnamefont {Felden}}, \bibinfo {author} {\bibfnamefont {C.}~\bibnamefont {Gilet}}, \bibinfo {author} {\bibfnamefont {R.}~\bibnamefont {Grout}}, \bibinfo {author} {\bibfnamefont {M.~H.}\ \bibnamefont {de~Frahan}}, \bibinfo {author} {\bibfnamefont {E.}~\bibnamefont {Motheau}}, \bibinfo {author} {\bibfnamefont {A.}~\bibnamefont {Nonaka}}, \bibinfo {author} {\bibfnamefont {L.}~\bibnamefont {Owen}},  \emph {et~al.},\ }\bibfield  {title} {\enquote {\bibinfo {title} {Pelelmex: an {AMR} low mach number reactive flow simulation code without level sub-cycling},}\ }\href@noop {} {\bibfield  {journal} {\bibinfo  {journal} {J. Open Source Softw.}\ }\textbf {\bibinfo {volume} {8}},\ \bibinfo {pages} {116674} (\bibinfo {year} {2023})}\BibitemShut {NoStop}%
\bibitem [{\citenamefont {Henry~de Frahan}\ \emph {et~al.}(2023)\citenamefont {Henry~de Frahan}, \citenamefont {Rood}, \citenamefont {Day}, \citenamefont {Sitaraman}, \citenamefont {Yellapantula}, \citenamefont {Perry}, \citenamefont {Grout}, \citenamefont {Almgren}, \citenamefont {Zhang}, \citenamefont {Bell} \emph {et~al.}}]{henry2023}%
  \BibitemOpen
  \bibfield  {author} {\bibinfo {author} {\bibfnamefont {M.~T.}\ \bibnamefont {Henry~de Frahan}}, \bibinfo {author} {\bibfnamefont {J.~S.}\ \bibnamefont {Rood}}, \bibinfo {author} {\bibfnamefont {M.~S.}\ \bibnamefont {Day}}, \bibinfo {author} {\bibfnamefont {H.}~\bibnamefont {Sitaraman}}, \bibinfo {author} {\bibfnamefont {S.}~\bibnamefont {Yellapantula}}, \bibinfo {author} {\bibfnamefont {B.~A.}\ \bibnamefont {Perry}}, \bibinfo {author} {\bibfnamefont {R.~W.}\ \bibnamefont {Grout}}, \bibinfo {author} {\bibfnamefont {A.}~\bibnamefont {Almgren}}, \bibinfo {author} {\bibfnamefont {W.}~\bibnamefont {Zhang}}, \bibinfo {author} {\bibfnamefont {J.~B.}\ \bibnamefont {Bell}},  \emph {et~al.},\ }\bibfield  {title} {\enquote {\bibinfo {title} {Pelec: An adaptive mesh refinement solver for compressible reacting flows},}\ }\href@noop {} {\bibfield  {journal} {\bibinfo  {journal} {Int. J. High Perform. Comput. Appl.}\ }\textbf {\bibinfo {volume} {37}},\ \bibinfo {pages} {115--131} (\bibinfo {year} {2023})}\BibitemShut
  {NoStop}%
\bibitem [{\citenamefont {An}\ \emph {et~al.}(2021)\citenamefont {An}, \citenamefont {Zhang}, \citenamefont {Zhang}, \citenamefont {Mao}, \citenamefont {Wei}, \citenamefont {Wang}, \citenamefont {Huang},\ and\ \citenamefont {Tan}}]{an2021}%
  \BibitemOpen
  \bibfield  {author} {\bibinfo {author} {\bibfnamefont {Z.}~\bibnamefont {An}}, \bibinfo {author} {\bibfnamefont {M.}~\bibnamefont {Zhang}}, \bibinfo {author} {\bibfnamefont {W.}~\bibnamefont {Zhang}}, \bibinfo {author} {\bibfnamefont {R.}~\bibnamefont {Mao}}, \bibinfo {author} {\bibfnamefont {X.}~\bibnamefont {Wei}}, \bibinfo {author} {\bibfnamefont {J.}~\bibnamefont {Wang}}, \bibinfo {author} {\bibfnamefont {Z.}~\bibnamefont {Huang}}, \ and\ \bibinfo {author} {\bibfnamefont {H.}~\bibnamefont {Tan}},\ }\bibfield  {title} {\enquote {\bibinfo {title} {Emission prediction and analysis on {CH4}/{NH3}/air swirl flames with {LES-FGM} method},}\ }\href@noop {} {\bibfield  {journal} {\bibinfo  {journal} {Fuel}\ }\textbf {\bibinfo {volume} {304}},\ \bibinfo {pages} {121370} (\bibinfo {year} {2021})}\BibitemShut {NoStop}%
\bibitem [{\citenamefont {Sweeney}\ \emph {et~al.}(2012{\natexlab{a}})\citenamefont {Sweeney}, \citenamefont {Hochgreb}, \citenamefont {Dunn},\ and\ \citenamefont {Barlow}}]{sweeney2012}%
  \BibitemOpen
  \bibfield  {author} {\bibinfo {author} {\bibfnamefont {M.~S.}\ \bibnamefont {Sweeney}}, \bibinfo {author} {\bibfnamefont {S.}~\bibnamefont {Hochgreb}}, \bibinfo {author} {\bibfnamefont {M.~J.}\ \bibnamefont {Dunn}}, \ and\ \bibinfo {author} {\bibfnamefont {R.~S.}\ \bibnamefont {Barlow}},\ }\bibfield  {title} {\enquote {\bibinfo {title} {The structure of turbulent stratified and premixed methane/air flames {II}: Swirling flows},}\ }\href@noop {} {\bibfield  {journal} {\bibinfo  {journal} {Combust. Flame}\ }\textbf {\bibinfo {volume} {159}},\ \bibinfo {pages} {2912--2929} (\bibinfo {year} {2012}{\natexlab{a}})}\BibitemShut {NoStop}%
\bibitem [{\citenamefont {Mao}\ \emph {et~al.}(2023)\citenamefont {Mao}, \citenamefont {Lin}, \citenamefont {Zhang}, \citenamefont {Zhang}, \citenamefont {Xu},\ and\ \citenamefont {Chen}}]{mao2023deep}%
  \BibitemOpen
  \bibfield  {author} {\bibinfo {author} {\bibfnamefont {R.}~\bibnamefont {Mao}}, \bibinfo {author} {\bibfnamefont {M.}~\bibnamefont {Lin}}, \bibinfo {author} {\bibfnamefont {Y.}~\bibnamefont {Zhang}}, \bibinfo {author} {\bibfnamefont {T.}~\bibnamefont {Zhang}}, \bibinfo {author} {\bibfnamefont {Z.-Q.~J.}\ \bibnamefont {Xu}}, \ and\ \bibinfo {author} {\bibfnamefont {Z.~X.}\ \bibnamefont {Chen}},\ }\bibfield  {title} {\enquote {\bibinfo {title} {Deepflame: A deep learning empowered open-source platform for reacting flow simulations},}\ }\href@noop {} {\bibfield  {journal} {\bibinfo  {journal} {Comput. Phys. Commun.}\ }\textbf {\bibinfo {volume} {291}},\ \bibinfo {pages} {108842} (\bibinfo {year} {2023})}\BibitemShut {NoStop}%
\bibitem [{\citenamefont {Weller}\ \emph {et~al.}(1998)\citenamefont {Weller}, \citenamefont {Tabor}, \citenamefont {Jasak},\ and\ \citenamefont {Fureby}}]{Weller1998}%
  \BibitemOpen
  \bibfield  {author} {\bibinfo {author} {\bibfnamefont {H.~G.}\ \bibnamefont {Weller}}, \bibinfo {author} {\bibfnamefont {G.}~\bibnamefont {Tabor}}, \bibinfo {author} {\bibfnamefont {H.}~\bibnamefont {Jasak}}, \ and\ \bibinfo {author} {\bibfnamefont {C.}~\bibnamefont {Fureby}},\ }\bibfield  {title} {\enquote {\bibinfo {title} {A tensorial approach to computational continuum mechanics using object-oriented techniques},}\ }\href@noop {} {\bibfield  {journal} {\bibinfo  {journal} {Comput. phys.}\ }\textbf {\bibinfo {volume} {6}},\ \bibinfo {pages} {620--631} (\bibinfo {year} {1998})}\BibitemShut {NoStop}%
\bibitem [{\citenamefont {Goodwin}(2022)}]{goodwin2002}%
  \BibitemOpen
  \bibfield  {author} {\bibinfo {author} {\bibfnamefont {D.~G.}\ \bibnamefont {Goodwin}},\ }\href@noop {} {\bibfield  {journal} {\bibinfo  {journal} {California Institute of Technology}\ }\textbf {\bibinfo {volume} {32}} (\bibinfo {year} {2022})}\BibitemShut {NoStop}%
\bibitem [{\citenamefont {Paszke}\ \emph {et~al.}(2019)\citenamefont {Paszke}, \citenamefont {Gross}, \citenamefont {Massa}, \citenamefont {Lerer}, \citenamefont {Bradbury}, \citenamefont {Chanan}, \citenamefont {Killeen}, \citenamefont {Lin}, \citenamefont {Gimelshein}, \citenamefont {Antiga} \emph {et~al.}}]{paszke2019}%
  \BibitemOpen
  \bibfield  {author} {\bibinfo {author} {\bibfnamefont {A.}~\bibnamefont {Paszke}}, \bibinfo {author} {\bibfnamefont {S.}~\bibnamefont {Gross}}, \bibinfo {author} {\bibfnamefont {F.}~\bibnamefont {Massa}}, \bibinfo {author} {\bibfnamefont {A.}~\bibnamefont {Lerer}}, \bibinfo {author} {\bibfnamefont {J.}~\bibnamefont {Bradbury}}, \bibinfo {author} {\bibfnamefont {G.}~\bibnamefont {Chanan}}, \bibinfo {author} {\bibfnamefont {T.}~\bibnamefont {Killeen}}, \bibinfo {author} {\bibfnamefont {Z.}~\bibnamefont {Lin}}, \bibinfo {author} {\bibfnamefont {N.}~\bibnamefont {Gimelshein}}, \bibinfo {author} {\bibfnamefont {L.}~\bibnamefont {Antiga}},  \emph {et~al.},\ }\bibfield  {title} {\enquote {\bibinfo {title} {Pytorch: An imperative style, high-performance deep learning library},}\ }\href@noop {} {\bibfield  {journal} {\bibinfo  {journal} {Adv Neural Inf Process Syst.}\ }\textbf {\bibinfo {volume} {32}} (\bibinfo {year} {2019})}\BibitemShut {NoStop}%
\bibitem [{\citenamefont {Naumov}\ \emph {et~al.}(2015)\citenamefont {Naumov}, \citenamefont {Arsaev}, \citenamefont {Castonguay}, \citenamefont {Cohen}, \citenamefont {Demouth}, \citenamefont {Eaton}, \citenamefont {Layton}, \citenamefont {Markovskiy}, \citenamefont {Reguly}, \citenamefont {Sakharnykh} \emph {et~al.}}]{naumov2015}%
  \BibitemOpen
  \bibfield  {author} {\bibinfo {author} {\bibfnamefont {M.}~\bibnamefont {Naumov}}, \bibinfo {author} {\bibfnamefont {M.}~\bibnamefont {Arsaev}}, \bibinfo {author} {\bibfnamefont {P.}~\bibnamefont {Castonguay}}, \bibinfo {author} {\bibfnamefont {J.}~\bibnamefont {Cohen}}, \bibinfo {author} {\bibfnamefont {J.}~\bibnamefont {Demouth}}, \bibinfo {author} {\bibfnamefont {J.}~\bibnamefont {Eaton}}, \bibinfo {author} {\bibfnamefont {S.}~\bibnamefont {Layton}}, \bibinfo {author} {\bibfnamefont {N.}~\bibnamefont {Markovskiy}}, \bibinfo {author} {\bibfnamefont {I.}~\bibnamefont {Reguly}}, \bibinfo {author} {\bibfnamefont {N.}~\bibnamefont {Sakharnykh}},  \emph {et~al.},\ }\bibfield  {title} {\enquote {\bibinfo {title} {Amgx: A library for {GPU} accelerated algebraic multigrid and preconditioned iterative methods},}\ }\href@noop {} {\bibfield  {journal} {\bibinfo  {journal} {SIAM J Sci Comput.}\ }\textbf {\bibinfo {volume} {37}},\ \bibinfo {pages} {S602--S626} (\bibinfo {year} {2015})}\BibitemShut {NoStop}%
\bibitem [{\citenamefont {Collobert}, \citenamefont {Bengio},\ and\ \citenamefont {Mari{\'e}thoz}(2002)}]{collobert2002}%
  \BibitemOpen
  \bibfield  {author} {\bibinfo {author} {\bibfnamefont {R.}~\bibnamefont {Collobert}}, \bibinfo {author} {\bibfnamefont {S.}~\bibnamefont {Bengio}}, \ and\ \bibinfo {author} {\bibfnamefont {J.}~\bibnamefont {Mari{\'e}thoz}},\ }\bibfield  {title} {\enquote {\bibinfo {title} {Torch: a modular machine learning software library},}\ }\href@noop {} {\bibfield  {journal} {\bibinfo  {journal} {Idiap}\ } (\bibinfo {year} {2002})}\BibitemShut {NoStop}%
\bibitem [{\citenamefont {DeepFlame}()}]{deepflame2024}%
  \BibitemOpen
  \bibfield  {author} {\bibinfo {author} {\bibnamefont {DeepFlame}},\ }\href@noop {} {\bibinfo  {journal} {https://github.com/deepmodeling/deepflame-dev}\ }\BibitemShut {NoStop}%
\bibitem [{\citenamefont {Kazakov}\ and\ \citenamefont {Frenklach}()}]{Kazakov}%
  \BibitemOpen
\bibfield  {journal} {  }\bibfield  {author} {\bibinfo {author} {\bibfnamefont {A.}~\bibnamefont {Kazakov}}\ and\ \bibinfo {author} {\bibfnamefont {M.}~\bibnamefont {Frenklach}},\ }\href@noop {} {\bibinfo  {journal} {http://www.me.berkeley.edu/drm/}\ }\BibitemShut {NoStop}%
\bibitem [{\citenamefont {Brown}, \citenamefont {Byrne},\ and\ \citenamefont {Hindmarsh}(1989)}]{brown1989}%
  \BibitemOpen
\bibfield  {journal} {  }\bibfield  {author} {\bibinfo {author} {\bibfnamefont {P.~N.}\ \bibnamefont {Brown}}, \bibinfo {author} {\bibfnamefont {G.~D.}\ \bibnamefont {Byrne}}, \ and\ \bibinfo {author} {\bibfnamefont {A.~C.}\ \bibnamefont {Hindmarsh}},\ }\bibfield  {title} {\enquote {\bibinfo {title} {Vode: A variable-coefficient ode solver},}\ }\href@noop {} {\bibfield  {journal} {\bibinfo  {journal} {SIAM J Sci Comput .}\ }\textbf {\bibinfo {volume} {10}},\ \bibinfo {pages} {1038--1051} (\bibinfo {year} {1989})}\BibitemShut {NoStop}%
\bibitem [{\citenamefont {Box}\ and\ \citenamefont {Cox}(1964)}]{Box1964}%
  \BibitemOpen
  \bibfield  {author} {\bibinfo {author} {\bibfnamefont {G.~E.}\ \bibnamefont {Box}}\ and\ \bibinfo {author} {\bibfnamefont {D.~R.}\ \bibnamefont {Cox}},\ }\bibfield  {title} {\enquote {\bibinfo {title} {An analysis of transformations},}\ }\href@noop {} {\bibfield  {journal} {\bibinfo  {journal} {J. Roy. Stat. Soc. B}\ }\textbf {\bibinfo {volume} {26}},\ \bibinfo {pages} {211--243} (\bibinfo {year} {1964})}\BibitemShut {NoStop}%
\bibitem [{\citenamefont {Zhang}\ \emph {et~al.}(2021{\natexlab{b}})\citenamefont {Zhang}, \citenamefont {An}, \citenamefont {Wei}, \citenamefont {Wang}, \citenamefont {Huang},\ and\ \citenamefont {Tan}}]{Zhang2021Emiss}%
  \BibitemOpen
  \bibfield  {author} {\bibinfo {author} {\bibfnamefont {M.}~\bibnamefont {Zhang}}, \bibinfo {author} {\bibfnamefont {Z.}~\bibnamefont {An}}, \bibinfo {author} {\bibfnamefont {X.}~\bibnamefont {Wei}}, \bibinfo {author} {\bibfnamefont {J.}~\bibnamefont {Wang}}, \bibinfo {author} {\bibfnamefont {Z.}~\bibnamefont {Huang}}, \ and\ \bibinfo {author} {\bibfnamefont {H.}~\bibnamefont {Tan}},\ }\bibfield  {title} {\enquote {\bibinfo {title} {Emission analysis of the {CH4}/{NH3}/air co-firing fuels in a model combustor},}\ }\href@noop {} {\bibfield  {journal} {\bibinfo  {journal} {Fuel}\ }\textbf {\bibinfo {volume} {291}},\ \bibinfo {pages} {120135} (\bibinfo {year} {2021}{\natexlab{b}})}\BibitemShut {NoStop}%
\bibitem [{\citenamefont {Zhang}\ \emph {et~al.}(2021{\natexlab{c}})\citenamefont {Zhang}, \citenamefont {Wei}, \citenamefont {Wang}, \citenamefont {Huang},\ and\ \citenamefont {Tan}}]{zhang2021blow}%
  \BibitemOpen
  \bibfield  {author} {\bibinfo {author} {\bibfnamefont {M.}~\bibnamefont {Zhang}}, \bibinfo {author} {\bibfnamefont {X.}~\bibnamefont {Wei}}, \bibinfo {author} {\bibfnamefont {J.}~\bibnamefont {Wang}}, \bibinfo {author} {\bibfnamefont {Z.}~\bibnamefont {Huang}}, \ and\ \bibinfo {author} {\bibfnamefont {H.}~\bibnamefont {Tan}},\ }\bibfield  {title} {\enquote {\bibinfo {title} {The blow-off and transient characteristics of co-firing ammonia/methane fuels in a swirl combustor},}\ }\href@noop {} {\bibfield  {journal} {\bibinfo  {journal} {Proc. Combust. Inst.}\ }\textbf {\bibinfo {volume} {38}},\ \bibinfo {pages} {5181--5190} (\bibinfo {year} {2021}{\natexlab{c}})}\BibitemShut {NoStop}%
\bibitem [{\citenamefont {Zhang}\ \emph {et~al.}(2021{\natexlab{d}})\citenamefont {Zhang}, \citenamefont {An}, \citenamefont {Wang}, \citenamefont {Wei}, \citenamefont {Jianayihan}, \citenamefont {Wang}, \citenamefont {Huang},\ and\ \citenamefont {Tan}}]{zhang2021regulation}%
  \BibitemOpen
  \bibfield  {author} {\bibinfo {author} {\bibfnamefont {M.}~\bibnamefont {Zhang}}, \bibinfo {author} {\bibfnamefont {Z.}~\bibnamefont {An}}, \bibinfo {author} {\bibfnamefont {L.}~\bibnamefont {Wang}}, \bibinfo {author} {\bibfnamefont {X.}~\bibnamefont {Wei}}, \bibinfo {author} {\bibfnamefont {B.}~\bibnamefont {Jianayihan}}, \bibinfo {author} {\bibfnamefont {J.}~\bibnamefont {Wang}}, \bibinfo {author} {\bibfnamefont {Z.}~\bibnamefont {Huang}}, \ and\ \bibinfo {author} {\bibfnamefont {H.}~\bibnamefont {Tan}},\ }\bibfield  {title} {\enquote {\bibinfo {title} {The regulation effect of methane and hydrogen on the emission characteristics of ammonia/air combustion in a model combustor},}\ }\href@noop {} {\bibfield  {journal} {\bibinfo  {journal} {Int. J. Hydrog. Energy}\ }\textbf {\bibinfo {volume} {4}},\ \bibinfo {pages} {21013--21025} (\bibinfo {year} {2021}{\natexlab{d}})}\BibitemShut {NoStop}%
\bibitem [{\citenamefont {Evans}\ \emph {et~al.}(2019)\citenamefont {Evans}, \citenamefont {Petre}, \citenamefont {Medwell},\ and\ \citenamefont {Parente}}]{evans2019}%
  \BibitemOpen
  \bibfield  {author} {\bibinfo {author} {\bibfnamefont {M.}~\bibnamefont {Evans}}, \bibinfo {author} {\bibfnamefont {C.}~\bibnamefont {Petre}}, \bibinfo {author} {\bibfnamefont {P.~R.}\ \bibnamefont {Medwell}}, \ and\ \bibinfo {author} {\bibfnamefont {A.}~\bibnamefont {Parente}},\ }\bibfield  {title} {\enquote {\bibinfo {title} {Generalisation of the eddy-dissipation concept for jet flames with low turbulence and low damk{\"o}hler number},}\ }\href@noop {} {\bibfield  {journal} {\bibinfo  {journal} {Proc. Combust. Inst.}\ }\textbf {\bibinfo {volume} {37}},\ \bibinfo {pages} {4497--4505} (\bibinfo {year} {2019})}\BibitemShut {NoStop}%
\bibitem [{\citenamefont {Chomiak}\ and\ \citenamefont {Karlsson}(1996)}]{chomiak1996}%
  \BibitemOpen
  \bibfield  {author} {\bibinfo {author} {\bibfnamefont {J.}~\bibnamefont {Chomiak}}\ and\ \bibinfo {author} {\bibfnamefont {A.}~\bibnamefont {Karlsson}},\ }\bibfield  {title} {\enquote {\bibinfo {title} {Flame liftoff in diesel sprays},}\ }\href@noop {} {\bibfield  {journal} {\bibinfo  {journal} {Symposium (International) on Combustion}\ }\textbf {\bibinfo {volume} {26}},\ \bibinfo {pages} {2557--2564} (\bibinfo {year} {1996})}\BibitemShut {NoStop}%
\bibitem [{\citenamefont {Kornev}\ and\ \citenamefont {Hassel}(2007)}]{kornev2007}%
  \BibitemOpen
  \bibfield  {author} {\bibinfo {author} {\bibfnamefont {N.}~\bibnamefont {Kornev}}\ and\ \bibinfo {author} {\bibfnamefont {E.}~\bibnamefont {Hassel}},\ }\bibfield  {title} {\enquote {\bibinfo {title} {Method of random spots for generation of synthetic inhomogeneous turbulent fields with prescribed autocorrelation functions},}\ }\href@noop {} {\bibfield  {journal} {\bibinfo  {journal} {Commun. Numer. Meth. Engng.}\ }\textbf {\bibinfo {volume} {23}},\ \bibinfo {pages} {35--43} (\bibinfo {year} {2007})}\BibitemShut {NoStop}%
\bibitem [{\citenamefont {Proch}\ \emph {et~al.}(2017)\citenamefont {Proch}, \citenamefont {Domingo}, \citenamefont {Vervisch},\ and\ \citenamefont {Kempf}}]{Proch2017}%
  \BibitemOpen
  \bibfield  {author} {\bibinfo {author} {\bibfnamefont {F.}~\bibnamefont {Proch}}, \bibinfo {author} {\bibfnamefont {P.}~\bibnamefont {Domingo}}, \bibinfo {author} {\bibfnamefont {L.}~\bibnamefont {Vervisch}}, \ and\ \bibinfo {author} {\bibfnamefont {A.~M.}\ \bibnamefont {Kempf}},\ }\bibfield  {title} {\enquote {\bibinfo {title} {Flame resolved simulation of a turbulent premixed bluff-body burner experiment. {P}art {I}: Analysis of the reaction zone dynamics with tabulated chemistry},}\ }\href@noop {} {\bibfield  {journal} {\bibinfo  {journal} {Combust. Flame}\ }\textbf {\bibinfo {volume} {180}},\ \bibinfo {pages} {321--339} (\bibinfo {year} {2017})}\BibitemShut {NoStop}%
\bibitem [{\citenamefont {Singh}, \citenamefont {Chander},\ and\ \citenamefont {Ray}(2012)}]{Singh2012}%
  \BibitemOpen
  \bibfield  {author} {\bibinfo {author} {\bibfnamefont {G.}~\bibnamefont {Singh}}, \bibinfo {author} {\bibfnamefont {S.}~\bibnamefont {Chander}}, \ and\ \bibinfo {author} {\bibfnamefont {A.}~\bibnamefont {Ray}},\ }\bibfield  {title} {\enquote {\bibinfo {title} {Heat transfer characteristics of natural gas/air swirling flame impinging on a flat surface},}\ }\href@noop {} {\bibfield  {journal} {\bibinfo  {journal} {Exp. Therm. Fluid Sci.}\ }\textbf {\bibinfo {volume} {41}},\ \bibinfo {pages} {165--176} (\bibinfo {year} {2012})}\BibitemShut {NoStop}%
\bibitem [{\citenamefont {Gaucherand}\ \emph {et~al.}(2023)\citenamefont {Gaucherand}, \citenamefont {Laera}, \citenamefont {Schulze-Netzer},\ and\ \citenamefont {Poinsot}}]{gaucherand2023}%
  \BibitemOpen
  \bibfield  {author} {\bibinfo {author} {\bibfnamefont {J.}~\bibnamefont {Gaucherand}}, \bibinfo {author} {\bibfnamefont {D.}~\bibnamefont {Laera}}, \bibinfo {author} {\bibfnamefont {C.}~\bibnamefont {Schulze-Netzer}}, \ and\ \bibinfo {author} {\bibfnamefont {T.}~\bibnamefont {Poinsot}},\ }\bibfield  {title} {\enquote {\bibinfo {title} {Intrinsic instabilities of hydrogen and hydrogen/ammonia premixed flames: Influence of equivalence ratio, fuel composition and pressure},}\ }\href@noop {} {\bibfield  {journal} {\bibinfo  {journal} {Combust. Flame}\ }\textbf {\bibinfo {volume} {256}},\ \bibinfo {pages} {112986} (\bibinfo {year} {2023})}\BibitemShut {NoStop}%
\bibitem [{\citenamefont {Sweeney}\ \emph {et~al.}(2012{\natexlab{b}})\citenamefont {Sweeney}, \citenamefont {Hochgreb}, \citenamefont {Dunn},\ and\ \citenamefont {Barlow}}]{sweeney2012struc}%
  \BibitemOpen
  \bibfield  {author} {\bibinfo {author} {\bibfnamefont {M.~S.}\ \bibnamefont {Sweeney}}, \bibinfo {author} {\bibfnamefont {S.}~\bibnamefont {Hochgreb}}, \bibinfo {author} {\bibfnamefont {M.~J.}\ \bibnamefont {Dunn}}, \ and\ \bibinfo {author} {\bibfnamefont {R.~S.}\ \bibnamefont {Barlow}},\ }\bibfield  {title} {\enquote {\bibinfo {title} {The structure of turbulent stratified and premixed methane/air flames {I}: Non-swirling flows},}\ }\href@noop {} {\bibfield  {journal} {\bibinfo  {journal} {Combust. Flame}\ }\textbf {\bibinfo {volume} {159}},\ \bibinfo {pages} {2896--2911} (\bibinfo {year} {2012}{\natexlab{b}})}\BibitemShut {NoStop}%
\bibitem [{\citenamefont {Qian}\ \emph {et~al.}(2022)\citenamefont {Qian}, \citenamefont {Zou}, \citenamefont {Lu},\ and\ \citenamefont {Yao}}]{qian2022}%
  \BibitemOpen
  \bibfield  {author} {\bibinfo {author} {\bibfnamefont {X.}~\bibnamefont {Qian}}, \bibinfo {author} {\bibfnamefont {C.}~\bibnamefont {Zou}}, \bibinfo {author} {\bibfnamefont {H.}~\bibnamefont {Lu}}, \ and\ \bibinfo {author} {\bibfnamefont {H.}~\bibnamefont {Yao}},\ }\bibfield  {title} {\enquote {\bibinfo {title} {Large-eddy simulation of {C}ambridge-{S}andia stratified flames under high swirl},}\ }\href@noop {} {\bibfield  {journal} {\bibinfo  {journal} {Combust. Flame}\ }\textbf {\bibinfo {volume} {244}},\ \bibinfo {pages} {112241} (\bibinfo {year} {2022})}\BibitemShut {NoStop}%
\bibitem [{\citenamefont {Turkeri}, \citenamefont {Zhao},\ and\ \citenamefont {Muradoglu}(2021)}]{turkeri2021}%
  \BibitemOpen
  \bibfield  {author} {\bibinfo {author} {\bibfnamefont {H.}~\bibnamefont {Turkeri}}, \bibinfo {author} {\bibfnamefont {X.}~\bibnamefont {Zhao}}, \ and\ \bibinfo {author} {\bibfnamefont {M.}~\bibnamefont {Muradoglu}},\ }\bibfield  {title} {\enquote {\bibinfo {title} {Large eddy simulation/probability density function modeling of turbulent swirling stratified flame series},}\ }\href@noop {} {\bibfield  {journal} {\bibinfo  {journal} {Phys. Fluids}\ }\textbf {\bibinfo {volume} {33}},\ \bibinfo {pages} {025117} (\bibinfo {year} {2021})}\BibitemShut {NoStop}%
\end{thebibliography}%

\end{document}